# Pretreatment MRI Reveals a Latent, Molecular-Subtype-Independent Structural Phenotype That Organises Treatment Trajectories and Recurrence Risk

**Authors:** Dattatreya Kantha[1], Murray H. Loew[1]
**Affiliation:** [1] Medical Imaging & Image Analysis Laboratory, Department of Biomedical Engineering, George Washington University, Science and Engineering Hall, 800 22nd St., NW, Suite 5000, Washington, DC 20052, USA
**Corresponding author:** Murray H. Loew, Ph.D., P.E. loew@gwu.edu
**First author contact:** Dattatreya Kantha, M.S. dattatreya.kantha@gwu.edu

## Abstract

Pathologic complete response and tumour shrinkage measure whether breast cancer responds to neoadjuvant therapy, but not whether that response was structurally favourable, persistent, or hidden beneath volume loss. We built an outcome-blind longitudinal DCE-MRI manifold from I-SPY2 trajectories to test whether pretreatment imaging carries a structural response phenotype missed by conventional descriptors. The dominant axis of response geometry was not recoverable from the full clinical and genomic stack — age, receptor subtype, MammaPrint, PAM50, treatment arm, and tumour burden — but became strongly recoverable once baseline structural entropy was added (cross-validated $R^2 = -0.023$ to 0.876; $\Delta R^2 = 0.899$). A constrained representation mapping recovered the same axes as unconstrained decomposition ($|r| \geq 0.952$; ARI = 1.000), establishing the structure as intrinsic rather than a post-hoc interpretation. The phenotype persisted through therapy (88.2% state retention; $\rho = 0.736$), and as treatment proceeded the volumetric signal faded while entropy stayed separated — a crossover from burden to structural persistence. Among complete responders, structurally disordered tumours could shrink more early yet remain structurally disordered ($\rho = -0.265$; size-adjusted $\rho = -0.262$), a volumetric deception invisible to endpoint labels. External analyses in UCSF, I-SPY1, and Duke established recurrence relevance under representation-dependent boundaries, and a representation-family commensurability assessment showed why feature-name matching is insufficient: the same label can fail, transport, or entangle with extraction geometry. Pretreatment MRI therefore exposes a structural response phenotype that endpoint-based language leaves invisible — including, among complete responders, a pretreatment imaging signal of structurally distinct response states that awaits prospective validation.

## Keywords

*DCE-MRI, longitudinal radiomics, manifold learning, structural entropy, breast cancer, radiomic transportability*

## 1. Introduction

Neoadjuvant therapy for breast cancer is judged by disappearance. Pathologic complete response — the absence of residual invasive disease at surgery — is the field's dominant endpoint [1, 2, 21], and volumetric shrinkage on serial imaging is its in-treatment proxy. Both ask the same question: how much tumour is left? Neither asks a deeper one: what was the structural biology of the response, and was it favourable, adverse, persistent, or merely concealed beneath the shrinkage? pCR is a surgical endpoint; volume is a burden measurement; and a tumour can lose volume without its underlying architecture becoming benign. The structural quality of a response is not its magnitude — yet our response frameworks measure only the latter.

This gap is sharpest where the endpoint looks most reassuring. A complete pathologic response can still compress different structural histories into the same favourable label: one tumour may enter therapy with organised, low-complexity architecture, while another may remain structurally disordered even as it shrinks. If pretreatment imaging encodes that distinction, the information is present before therapy begins but is not being read. This matters clinically because endpoint-equivalent patients — especially those classified as complete responders — may still differ in the structural biology relevant to recurrence risk, surveillance, and risk-stratification decisions.

Dynamic contrast-enhanced MRI is well suited to look for it. Acquired serially across treatment, DCE-MRI captures not only tumour burden but the spatial heterogeneity of enhancement — the structural texture of a lesion and its evolution over time [3, 4, 11, 17]. Most radiomic work, however, points this signal at a fixed target, training models to predict pCR or recurrence from imaging features [5, 6, 7, 9, 15]. Prediction is useful, but it presupposes we already know what to look for. It does not ask whether the trajectories contain an intrinsic geometry of response — a latent structure organising how tumours move through treatment, independent of any outcome label. Discovering that geometry, rather than predicting a known endpoint, is the task we take up here.

We therefore asked three questions. First, does pretreatment DCE-MRI reveal an unsupervised structural response phenotype — a latent organisation of treatment trajectories recoverable without any outcome supervision? Second, is that phenotype independent of the descriptors oncology already uses — molecular subtype, PAM50, MammaPrint genomic risk, treatment arm, age, and tumour burden — or merely a reflection of them? Third, under what conditions can such a phenotype be trusted in an external cohort?

The third question is less routine than it sounds. External validation is usually treated as a question about cohorts: does the result replicate in new patients? For radiomic features it is first a question about representation. A feature named "entropy" is not a fixed biological quantity; it is the output of an extraction pipeline, and the same name can denote different measurements when computed from contour-defined enhancement maps in one cohort, bounding-box-derived features in another, or volume-only trajectory projections in a third. Two cohorts can share a feature's name, its software, and its acquisition lineage and still not share its biology. Validating a phenotype across cohorts therefore requires auditing whether the feature carrying it remains commensurable — a representation-family question that precedes, and can override, the survival analysis itself.

This study makes seven contributions. (1) We identify an entropy-organised structural phenotype that governs the geometry of longitudinal treatment response, recovered without outcome supervision. (2) We show that the dominant structural axis is not recoverable from the conventional clinical and genomic descriptor stack, but becomes recoverable when baseline entropy is added. (3) We demonstrate that it persists through therapy as an individual-level property, not a transient baseline feature. (4) We characterise an information crossover, in which volumetric separation between phenotype groups decays across treatment while structural separation is maintained. (5) We expose a within-pCR volumetric deception, in which structurally disordered tumours can shrink fastest yet remain structurally disordered, invisible to endpoint labels. (6) We test, across three external cohorts — UCSF, I-SPY1, and Duke — where the phenotype carries recurrence relevance and where representation boundaries limit inference. (7) We develop and apply a representation-family commensurability assessment (RFCA) as reusable validation infrastructure, formalising why feature-name matching is insufficient for radiomic transport and how to audit it.

Throughout, discovery and model fitting are confined to a locked training partition; every external cohort and the held-out test set is evaluated under pre-specified, training-derived transforms — so the structural phenotype is discovered once and tested, never tuned. Together, these analyses recast pretreatment MRI from a response-prediction input into a structural readout of response biology that endpoint-based language leaves unresolved.

## 2. Methods

### 2.1. Study Design and Inference Contract

This study follows a partition-locked modular design in which biological characterisation, outcome prediction, and external validation are addressed through sequentially constructed, logically independent components. A complete operational summary is provided in the Supplementary Reproducibility Checklist (Table S14.1). The I-SPY2 cohort was partitioned into training and test sets before any feature scaling, dimensionality reduction, clustering, or predictive modelling. All scaling parameters, PCA bases, cluster centroids, model coefficients, calibration parameters, and decision thresholds were learned exclusively from the training partition; the held-out test set was evaluated only after all modelling decisions were finalized. No outcome labels were used during manifold construction.

Three complementary external validation arms were evaluated under two distinct inference modes. For UCSF, subjects were projected through the pre-specified, training-derived I-SPY2 manifold with no retraining or recalibration. For Duke, the locked I-SPY2 training-partition entropy median (2.6500518) was transferred directly as a stratification threshold without any Duke-side parameter estimation, necessitated by the pre-specified representation-family audit (§2.10). For I-SPY1, subjects were projected through pre-specified Tier-2 FTV-only transforms. These protocols are complementary but logically independent: UCSF validates trajectory-space manifold geometry; Duke validates the transportable structural-heterogeneity core against long-term recurrence; I-SPY1 demonstrates within-cohort feature-family specificity of external transport. The I-SPY2 cohort was split 70/30 (stratified on pCR, seed = 42), yielding 494 training subjects (pCR rate 31.8%) and 212 held-out test subjects (pCR rate 32.1%). Full pipeline specifications are in Supplementary Section S1.2.

## 2.2. Datasets

### 2.2.1. Discovery Cohort — I-SPY2

The internal development cohort comprised 719 subjects from the TCIA I-SPY2 Imaging Cohort 1 (ACRIN 6698) [18, 20], a prospective adaptive neoadjuvant chemotherapy trial across 22 academic centers [19]. Subjects underwent serial DCE-MRI at four protocol-defined timepoints (T0–T3) with expert-derived tumour segmentations throughout. The primary endpoint was pathologic complete response (pCR; overall rate 32.7%). Scanner consistency was high (manufacturer 99.4%; field strength 99.9%). Full preprocessing details are in Supplementary Section S1.2.

### 2.2.2. External Validation Cohort — UCSF Breast-MRI-NACT-Pilot

External trajectory validation was performed in the TCIA Breast-MRI-NACT-Pilot cohort [56], acquired at UCSF between 1995 and 2002 on GE 1.5T systems. Of 64 identified subjects, 49 had complete T0–T2 trajectories; 15 were excluded for missing baseline-to-mid-treatment imaging before outcome analysis. The primary endpoint was recurrence-free survival (15 events, median follow-up 61 months). Baseline tumour burden differed approximately four-fold from I-SPY2 training (median FTV 20.7 cc versus 5.0 cc). Full preprocessing details are in Supplementary Section S1.2.2.

### 2.2.3. External Validation Cohort — Duke Breast Cancer MRI

The Duke Breast Cancer MRI dataset [57] is a publicly available single-centre cohort organised across three patient-level data layers: a clinical outcomes table, a precomputed imaging-feature matrix (922 subjects × 530 features), and a tumour-localization annotation table. The primary endpoint was distant recurrence-free survival (DRFS), defined by a numeric value in the days-to-distant-recurrence field; the generic binary recurrence flag was not used because it conflates all recurrence types. After excluding 13 subjects with unresolvable endpoint ambiguity and 1 subject with non-positive follow-up, the DRFS-eligible cohort comprised 908 subjects with 76 events across four subtype groups (Table 1). The primary imaging variable was Entropy_Tumor, identified by the pre-specified representation-family audit (§2.10) as the sole Duke entropy representation occupying the same numeric regime as the I-SPY2 training reference (Cohen's $d = -0.020$). The locked I-SPY2 training-partition threshold (2.6500518) was applied without Duke-side estimation. Full endpoint derivation is in Supplementary Section S8.1.

### 2.2.4. External Validation Cohort — I-SPY1 (ACRIN 6657)

The I-SPY1 dataset (TCIA ACRIN 6657) [12, 63] provided an additional external Tier-2 FTV validation cohort. Tumour-volume extraction from archived DICOM SEG objects produced a projection-eligible cohort of 126 subjects with complete T0/T1/T2 extracted FTV and 39 recurrence events. The I-SPY1 arm serves two methodological roles: demonstrating within-cohort feature-family specificity of external transport under the RFCA framework, and assessing whether pre-specified Tier-2 FTV-only manifold geometry transports coherently into a structurally independent acquisition-era cohort. All I-SPY1 analyses used the same pre-specified Tier-2 artifacts used for UCSF projection with no I-SPY1-derived parameter estimation. Full cohort construction details are in Supplementary Section S9.1.

**Table 1. Updated Four-Cohort Study Design Characteristics**

| Characteristic | I-SPY2 | UCSF | I-SPY1 | Duke |
|---|---|---|---|---|
| **Role** | Discovery | External validation (trajectory) | External validation (feature-family specificity; geometry transport) | External validation (heterogeneity axis) |
| **Acquisition era** | 2010–2016 | 1995–2002 | 1983–2001 (ACRIN 6657) | 2000s–2010s |
| **Scanner environment** | Multi-center, mixed 1.5T/3T | Single-center, GE 1.5T | Multi-site, mixed GE scanners | Single-center, mixed scanner vendors and field strengths (1.5T/3T) |
| **Subjects (analysis-eligible)** | 539 trajectory-complete (TRAIN 373, TEST 166) | 49 | 126 (T0/T1/T2 complete FTV) | 908 (DRFS-eligible) |
| **Primary endpoint** | pCR (standardized) | RFS (15 events) | RFS (39 events); geometry transport (primary) | DRFS (76 events) |
| **Imaging representation** | Longitudinal DCE-MRI trajectories (T0–T3) | Longitudinal DCE-MRI trajectories (T0–T2) | Longitudinal DCE-MRI trajectories (T0–T2, extracted FTV) | Precomputed baseline imaging features (530 columns) |
| **Entropy representation** | Entropy_T0 (PyRadiomics, contour-based) | Entropy_T0 (PyRadiomics, PE1-phase); structurally disjoint, not projected (FTV-only) | PARTIALLY_SHIFTED (d = 1.354; entropy not projected) | Entropy_Tumor (precomputed, bounding-box-based) |
| **Molecular subtype distribution** | TNBC 39%, HR+/HER2− 38%, HR+/HER2+ 15%, HR−/HER2+ 8% | Not available | Available (HR status) | HR+/HER2− 65%, TNBC 18%, HR+/HER2+ 11%, HR−/HER2+ 6% |
| **Median baseline FTV / tumor entropy** | 5.0 cc (TRAIN) | 20.7 cc | 18.5 cc (State 1) | Entropy_Tumor mean 2.034 |
| **Validation protocol** | Partition-locked inference contract | Locked manifold projection (FTV-only Tier 2) | Locked Tier-2 FTV-only projection (identical artifacts to UCSF) | Locked threshold-transfer (I-SPY2 train median) |
| **External validation question** | — | Does manifold geometry generalize under acquisition shift? | Does FTV dynamics pass Stage 1 where entropy fails in the same cohort? Does pre-specified geometry transport coherently across acquisition eras? | Does the heterogeneity axis carry recurrence-risk relevance after clinicopathologic adjustment? |

### 2.3. Feature Extraction, Longitudinal Representation, and Cross-Cohort Transportability Assessment

Two complementary imaging features were extracted at each protocol timepoint: functional tumour volume (FTV) and structural entropy. FTV was computed as the physical volume of the expert-defined tumour ROI in cubic centimetres from native voxel geometry; it captures tumour burden — how much disease is present. Structural entropy, by contrast, captures how internally disordered that disease is, and was selected as the primary heterogeneity descriptor because it summarises intratumoural architectural complexity as a single interpretable scalar, computed reproducibly at each timepoint without requiring voxel-level spatial correspondence across visits. Prior work has independently linked baseline entropy to treatment resistance and adverse prognosis in breast cancer [13, 14]. Structural entropy was extracted from SER signal-enhancement-ratio maps using PyRadiomics [22] (original_firstorder_Entropy, bin width 0.1) in accordance with IBSI [23] first-order entropy definitions; full configuration is in Supplementary Table S1.1 [24]. Lower entropy reflects a more spatially organised enhancement pattern, whereas higher entropy reflects greater intratumoural architectural heterogeneity. In this setting, first-order enhancement-intensity heterogeneity is used as a tumour-level imaging proxy for architectural and vascular disorganisation. We use "structural entropy" for this measured scalar feature, "structural axis" for the latent manifold dimension it dominates, and "structural phenotype" for the patient-level position along that axis; the structural axis therefore separates tumours primarily by baseline architectural organisation rather than by tumour size. The orthogonality of FTV and entropy was confirmed by consistently low Spearman correlations at each timepoint (ρ range 0.05–0.10).

Longitudinal representations comprised cumulative and step delta features for FTV and entropy. A training-only restriction yielded 10 stable, size-independent baseline texture features as structural context at T0 (Supplementary Table S1.2). Eligibility was determined by data availability without imputation: 670 subjects were M3-eligible (T0–T1), 620 M4-eligible (T0–T2), and 539 trajectory-complete (T0–T3). Prior to external manifold projection, entropy distributions were compared between I-SPY2 and UCSF: although both used the same PyRadiomics first-order feature and fixed bin-width discretization, they differed in source-image representation (I-SPY2 SER-native; UCSF PE1-phase), and the resulting distributions were structurally disjoint (I-SPY2: mean 2.04; UCSF: mean 6.5). Entropy was therefore excluded from cross-cohort projection and a reduced FTV-only feature set was used for UCSF.

### 2.4. Latent Dynamic Response State (LDRS) Manifold Construction

The LDRS framework constructs an unsupervised low-dimensional manifold of longitudinal imaging dynamics from the training cohort without outcome labels at any stage. It is not designed as a pCR predictor; rather, it is a structural-dynamic response descriptor encoding how tumour phenotype and burden evolve under therapy. This architectural separation — unsupervised biological characterisation followed by locked inference — enables each component to be independently audited, validated, and applied across cohorts.

LDRS construction required trajectory-complete subjects ($n = 539$; training $n = 373$, test $n = 166$). Each subject was represented by a 24-dimensional feature vector comprising raw longitudinal values (FTV and entropy at T0–T3; 8 features), cumulative delta features relative to baseline (6 features), and 10 restricted baseline texture features at T0. A StandardScaler and PCA were fit exclusively on the training partition (10 components, 90.15% cumulative variance), with training-only percentile clipping applied for geometric stabilization. Cluster number was selected from pre-specified candidates $\{3, 4, 5\}$ by bootstrap ARI stability on training only ($k = 3$: median ARI = 0.980 versus 0.898 for $k = 4$ and 0.853 for $k = 5$); k-means yielded cluster sizes of 14, 205, and 154 (silhouette score 0.435). Test subjects were assigned by locked nearest-centroid classification. For UCSF, a separate FTV-only scaler and PCA were trained on I-SPY2 training subjects under the same inference contract because entropy was non-transportable across acquisition eras (§2.3). Full specifications are in Supplementary Sections S2.1–S2.2.

Manifold axes were interpreted using train-only Spearman correlations: PC1 aligned with baseline structural entropy ($\rho = -0.933$); PC2 aligned with baseline FTV ($\rho = 0.809$); and PC3 aligned with therapy-induced entropy change ($\rho = 0.679$ with ΔEntropy_30). Cluster labels were anchored post hoc using training-set trajectory summaries. A per-subject confidence score was constructed from three training-derived components — bootstrap stability, centroid alignment, and temporal consistency — with weights fixed prior to outcome evaluation. Full details are in Supplementary Sections S2.3–S2.4.

#### 2.4.1. Biologically Constrained Manifold Validation (DLRM)

Standard unsupervised manifold methods recover axes that can be interpreted biologically post hoc but provide no formal test of whether that interpretation is intrinsic to the data or incidental to the dimensionality reduction method. The Discriminative Latent Response Manifold (DLRM) addresses this by imposing the hypothesized axis structure as a construction constraint and using convergence between constrained and unconstrained solutions as a formal test of intrinsicness.

DLRM was constructed using sequential biologically anchored partial least squares [64] on the training partition exclusively. Axis 1 was anchored to baseline structural heterogeneity (Entropy_T0), axis 2 to baseline tumour burden (FTV_T0), and axis 3 to treatment-induced structural remodelling (ΔEntropy_30), with sequential deflation enforcing orthogonality. No clinical outcome labels were used. Agreement between DLRM and PCA was quantified by Pearson correlation between projected score vectors and cosine similarity between feature weight vectors. The integration criterion required score correlations $\geq 0.95$ on axes 1 and 2. Results are reported in Section 3.2.3.

### 2.5. Biological Grounding and Molecular Independence Analyses

A pre-specified biological grounding analysis was performed exclusively on the training partition. Principal component axes were characterised using train-only Spearman correlations; cluster separation was evaluated using

Kruskal–Wallis tests with eta-squared effect sizes; and subtype independence was evaluated using chi-square testing with Cramér's V. No outcome labels were used. Within-subtype structural organisation was evaluated on the full trajectory-complete cohort (n = 539) as confirmatory characterisation (Supplementary Section S2.7).

Molecular independence was additionally assessed against two transcriptomic classifiers. MammaPrint status was evaluated using chi-square testing with Cramér's V and within-stratum Mann–Whitney U testing. PAM50 intrinsic subtypes were derived from the public I-SPY2-990 pretreatment transcriptomic resource (GSE194040 [58]; n = 987, 704 with matched imaging), with association assessed by chi-square and continuous independence by correlating Entropy_T0 with each PAM50 centroid score. Burden independence was tested by residualizing baseline entropy against baseline FTV and reconstructing PCA on the baseline feature family under the same training-only preprocessing configuration; results are in Section 3.2.1 and Supplementary Section S10.

### 2.6. Incremental pCR Prediction Ladder

A six-model elastic-net regularized logistic regression ladder was constructed to quantify the incremental predictive contribution of each evidence tier. Models were defined cumulatively: M0 anchored at enrollment demographics; M1 added molecular susceptibility markers (HR, HER2, MammaPrint); M2 added baseline imaging burden and structural phenotype; M3 added early response delta features at T1; M4 added locked LDRS state assignment; and M5 added full-course imaging summaries through T3. Supplementary Table S13.7 defines each model, evidence block, and eligible cohort sizes.

All models used elastic-net logistic regression with L1-to-L2 ratio and regularization strength C selected by cross-validation on the eligible training subset only. Each model was refit on the full eligible training subset and evaluated exactly once on the locked held-out test partition. AUROC, AUPRC, and Brier score were computed with bootstrap 95% confidence intervals from 2,000 stratified resamples. Probabilities were calibrated using isotonic regression fit exclusively on training data. Full configuration is in Supplementary Section S3.

### 2.7. Decision Zones

Decision thresholds were derived exclusively from out-of-fold training predictions. A rule-out threshold (t_out) targeted 90% sensitivity for non-pCR detection; a rule-in threshold (t_in) targeted 90% specificity for pCR detection. Zone assignment used an interval-aware rule: rule-out if p_hi < t_out; rule-in if p_lo > t_in; indeterminate if the interval spanned one or both thresholds. Both thresholds were locked after training-only derivation. Full analyses are in Supplementary Sections S4–S5.

### 2.8. Statistical Analysis

Model discrimination was assessed using AUROC and AUPRC with bootstrap 95% confidence intervals from 2,000 stratified resamples [30]. Calibration was assessed using Brier score [32], expected calibration error, and calibration slope. Incremental ladder comparisons used paired bootstrap ΔAUROC on intersection cohorts; DeLong paired tests [33] with false discovery rate correction [36] were performed as a complementary parametric assessment. Decision curve analysis followed the Vickers–Elkin formulation [34] for M2, M4, and M5 across the threshold range 0.05–0.50. A comprehensive robustness battery is reported in Supplementary Sections S4–S5. Latent states were derived by k-means clustering [26] and assessed with silhouette analysis [27] and the adjusted Rand index [28]; regularised models used the elastic net [29] with probability calibration [31]; survival analyses used Cox proportional-hazards regression [37], Kaplan–Meier estimation [38], permutation testing [39], and Harrell's concordance index [61]; cross-cohort harmonisation methods such as ComBat were referenced [40, 41]; and machine-learning models were implemented in scikit-learn [66], XGBoost [67], and support-vector machines [68].

All reported results satisfy pre-specified leakage safeguards. The train-test split was fixed before any modelling; all scaling and encoding steps were fit on training data only; thresholds were derived from training out-of-fold predictions; and the held-out test set was evaluated exactly once. A complete safeguards inventory is in Supplementary Table S14.1.

A pre-specified subtype-adjusted signal test evaluated whether pre-specified manifold variables contributed pCR discrimination beyond an explicit molecular baseline (HR + HER2 + MammaPrint) in the trajectory-complete cohort (training n = 373, test n = 166). Net reclassification improvement [35] for key ladder transitions is in Supplementary Section S5.5. Results of the subtype-adjusted signal test are in Section 3.3.2 and Supplementary Section S5.11.

### 2.9. UCSF External Validation Protocol

The UCSF cohort was processed independently of all I-SPY2 modelling decisions. Entropy non-transportability was confirmed as described in §2.3, activating the pre-specified Tier-2 reduced feature strategy. No ComBat correction or cohort-specific rescaling was applied. A reduced FTV-only manifold was constructed exclusively from I-SPY2 training subjects with complete T0–T2 trajectories (n = 440) using five FTV features, two-component PCA (81.8% variance), and k = 3 k-means. All transforms were serialized before any UCSF projection.

The 49 UCSF subjects were projected through locked transforms without recalibration. Because State 2 contained only two subjects after projection, the primary comparison was pre-specified as State 1 versus Non-State 1. Survival analysis used log-rank testing, ridge-penalized Cox regression, bootstrap stability (2,000 resamples), permutation testing (1,000 resamples), and leave-one-out influence analysis. A pre-specified anthracycline-only sensitivity analysis was also performed. Full specifications are in Supplementary Sections S6 and S6.7.

### 2.10. Representation-Family Commensurability Assessment (RFCA)

Radiomic features sharing a label class may be computed from fundamentally different source image representations across cohorts, producing values in incompatible numeric regimes despite identical extraction software and parameter settings — a mismatch invisible to feature-label matching and unaddressed by any current validation standard [8, 10, 23, 25]. The RFCA is a four-stage pre-specified protocol that determines representation-family commensurability before any outcome-associated analysis is performed, providing reusable methodological infrastructure for cross-cohort radiomic validation (Supplementary Fig. S7.1). Its threshold operationalizations follow established effect-size conventions [59] and are scoped as proof-of-concept implementations requiring prospective validation before adoption as fixed decision rules.

Because radiomic label classes can conceal incompatible numeric regimes, RFCA first audits feasibility: when too few discovery-cohort components match the external feature matrix for a composite to be projected — as occurred for the eleven-feature baseline heterogeneity index against Duke, where only one component matched — the assessment proceeds on the individual representations sharing the discovery anchor's label class (Supplementary Section S8.2). Each candidate is then carried through four pre-specified stages: Stage 1 tests numeric-regime compatibility against the discovery distribution (Cohen's d, with a shape-sensitive complement); Stage 2 diagnoses coupling to extraction geometry and pre-registers the covariate structure for outcome analysis; Stage 3 transfers the pre-specified discovery threshold and performs outcome analysis only under the compatible representation; and Stage 4 audits whether any surviving association remains biologically interpretable once extraction-geometry effects are removed, classifying a feature as geometry-entangled when its association is abolished under extent adjustment across the representation family. Full thresholds and procedural specifications are provided in Supplementary Section S7.1.

### 2.11. Duke External Validation — Distant Recurrence-Free Survival Analysis

All Duke DRFS analyses used the DRFS-eligible cohort (n = 908, 76 events). A binary stratification variable was defined by applying the pre-specified I-SPY2 training-partition median (2.6500518) to Duke Entropy_Tumor before any outcome access. The primary analysis used Kaplan–Meier estimation, log-rank testing, and unadjusted Cox regression, with Schoenfeld residual verification [60]. The primary adjusted model used reliably available receptor covariates (Entropy_high with HR and HER2 status; EPV = 25.3). A subtype-adjusted Cox model (EPV = 19.0) and a broader complete-case clinicopathologic model additionally including Nottingham grade and nodal stage (n = 611, 49 events; reduced to an events-per-variable ratio of approximately 5.4 by 31% grade missingness; Supplementary Section S8.1) were treated as supporting and sensitivity analyses, respectively. Full model results are in Supplementary Sections S8.5–S8.9. Four pre-specified sensitivity analyses — including Duke internal median threshold, subtype-stratified Cox regression, pCR discrimination in the NAT-eligible subset, and major-axis sensitivity analysis (Supplementary Section S8.11) — are reported in the corresponding results sections.

### 2.12. Longitudinal Trajectory Biology Analyses

Four trajectory analyses were performed in the I-SPY2 longitudinal analysis cohort (n = 706; subjects with baseline entropy and pCR labels, distinct from the strict T0–T3 LDRS-eligible subset used for manifold characterisation, n = 539) using the pre-specified Entropy_high/Entropy_low stratification (threshold 2.6500518), without outcome-informed threshold selection. Overall trajectory analysis compared FTV and entropy between groups at each timepoint and for cumulative delta features using Mann–Whitney U testing. Within-subtype analysis repeated these comparisons within each of the four molecular subtype groups. Within-pCR analysis repeated these comparisons within the pCR-positive subgroup — a null FTV result with a significant entropy result within this subgroup is direct evidence that entropy captures structural biology that pCR does not resolve. LDRS concordance analysis cross-tabulated the pre-specified entropy phenotype against unsupervised LDRS state assignment. Full trajectory visualizations are in Supplementary Section S11.

### 2.13. Structural Phenotype Characterisation Analyses

#### 2.13.1. Variance Decomposition of the Dominant Manifold Axis

To quantify whether the dominant manifold axis (PC1) could be recovered from conventional clinical and genomic descriptors, two linear models were compared using the locked training partition (n = 373). Model A included the full conventional baseline stack: baseline FTV, age, receptor subtype (HR/HER2), MammaPrint genomic risk, treatment arm, and PAM50 intrinsic subtype. Model B added baseline Shannon entropy to the same stack. Both models were evaluated using five-fold cross-validated $R^2$ to prevent overfitting. The incremental $R^2$ attributable to entropy quantifies the information content of the structural phenotype that is invisible to the conventional classification architecture.

#### 2.13.2. Structural Persistence Through Therapy

Persistence of the entropy-defined phenotype from baseline (T0) to post-treatment (T3) was evaluated at three complementary levels: continuous rank-order correlation (Spearman rho and Pearson r between T0 and T3 entropy values); binary state persistence (proportion of subjects retaining their median-split structural state, with a 2x2 transition matrix); and ordinal persistence (proportion of subjects remaining in the same or adjacent entropy quartile from T0 to T3). Mean signed and absolute entropy changes were computed to distinguish systematic directional drift from bidirectional perturbation. Extreme-state retention rates were computed for the bottom and top quartiles separately.

#### 2.13.3. Within-pCR Volumetric Deception Analysis

Within the pCR-positive population, the association between baseline structural entropy and early volumetric response was evaluated using Spearman rank correlation between Entropy_T0 and three early response measures: percent-change FTV (T0 to T1), log-ratio FTV, and raw delta FTV. Quartile-based contrasts compared the highest-entropy quartile against the lowest. To rule out a baseline-size confound, all volumetric response measures were residualised against FTV_T0 before recomputing entropy correlations. The analysis used the large-denominator pCR pool (n = 225 with entropy and FTV available; of these, 220 had longitudinal entropy-subgroup assignment and 217 had early percent-volume-change data) rather than the trajectory-complete subset.

#### 2.13.4. Support-Conditioned Continuous Manifold Risk

Continuous manifold geometry was evaluated against recurrence-free survival in the I-SPY1 external cohort using Cox proportional hazards regression with PC1 (standardised) as the predictor. Two analyses were compared: overall (all 126 projected subjects) and support-conditioned (restricted to n = 110 subjects within the 95th-percentile training support region, Within95Support = TRUE). A geometry battery additionally evaluated AssignedCentroidDistance and DistanceToTrainCenter as alternative continuous risk features. Distributional separation between in-support and out-of-support subjects was quantified using Cohen d across all geometric features. A disease-relevant HER2-positive subgroup analysis was pre-specified within the in-support subset.

#### 2.13.5. PC3 Remodeling Axis Characterization

To determine whether PC3 represents a biologically meaningful remodelling dimension or baseline leakage, incremental linear models were constructed for each entropy-change target (ΔEntropy_10, ΔEntropy_20, ΔEntropy_30). Models compared PC1-only, PC3-only, and PC1+PC3 specifications. To formally exclude baseline leakage, all entropy-change targets were residualised against Entropy_T0, FTV_T0, and both simultaneously before repeating the incremental comparison. PC3 was additionally tested as an incremental predictor of pCR in logistic regression to confirm that it does not function as an endpoint-discrimination axis.

## 3. Results

Sections 3.2–3.3 establish the entropy-organised response structure entirely within I-SPY2 and do not depend on external replication; their central result is that the dominant manifold axis is irreducible to the full conventional clinical and genomic stack yet recoverable from baseline entropy alone (Fig. 3). Later sections evaluate transportability, prognostic extension, and boundary conditions under pre-specified, training-derived evaluation protocols.

### 3.1. Cohort Characteristics

Table 2 summarizes characteristics of the LDRS-eligible cohorts. Within the I-SPY2 trajectory-complete subset, molecular subtype distributions were well balanced across training (n = 373) and test (n = 166) partitions ($\chi^2$ $p > 0.90$), confirming preserved biological composition between development and evaluation cohorts. Median baseline FTV differed approximately four-fold between I-SPY2 training and UCSF (5.0 cc versus 20.7 cc; KS $p = 5.0\times10^{-8}$; Cohen's $d = -0.35$), representing a substantial distributional shift treated as a deliberate stress test of locked-manifold generalisation under severe acquisition-era shift. UCSF recurrence analyses were performed under locked projection-only inference without recalibration or parameter refitting.

**Table 2. LDRS-Eligible Cohort Characteristics**

| Characteristic | I-SPY2 Train (n = 373) | I-SPY2 Test (n = 166) | UCSF (n = 49) |
|---|---|---|---|
| Median age, years | 49 | 50 | 52 |
| TNBC, n (%) | 143 (38.3) | 55 (33.1) | 11 (22) |
| HR+/HER2−, n (%) | 145 (38.9) | 68 (41.0) | 18 (37) |
| HR+/HER2+, n (%) | 54 (14.5) | 34 (20.5) | 6 (12) |
| HR−/HER2+, n (%) | 31 (8.3) | 9 (5.4) | 4 (8) |
| Unknown / unavailable subtype, n (%) | — | — | 10 (20) |
| pCR rate, n (%)† | 115 (30.8) | 51 (30.7) | 6/49 (12, proxy) |
| Median baseline FTV, cc | 5.0 | 5.3 | 20.7 |
| Recurrence events | — | — | 15 |
| Median follow-up, months | — | — | 61 |
| Acquisition era | 2010–2016 | 2010–2016 | 1995–2002 |
| Scanner field strength | 1.5T/3T | 1.5T/3T | 1.5T |

*†pCR rate in trajectory-complete subjects (30.8% training; 30.7% test) is comparable to the full pCR-modelling cohort rate (31.8% training; 32.1% test), indicating that trajectory-complete filtering did not materially bias the cohort toward responders. UCSF subtype annotations were unavailable for 10/49 subjects; percentages for UCSF subtype rows therefore reflect the full analysis-eligible cohort denominator.*

### 3.2. Unsupervised Manifold Structure and Biological Validation

*All analyses in this section were performed on the training partition exclusively; no pCR or survival labels were used at any stage.*

#### 3.2.1. Dominant Manifold Axes

When the LDRS manifold was constructed from 24-dimensional longitudinal DCE-MRI trajectories in 373 I-SPY2 training subjects without outcome supervision, the dominant axis of the learned geometry aligned with baseline structural heterogeneity — not volumetric response, not molecular subtype. PC1 showed a strong inverse correlation with pretreatment structural entropy (ρ = −0.933), establishing baseline architectural complexity as the primary organising axis. PC2 correlated strongly with baseline FTV (ρ = 0.809), capturing tumour burden as the secondary axis. PC3 correlated moderately with therapy-induced entropy change (ρ = 0.679 with ΔEntropy_30), representing treatment-induced structural reorganisation as the third and most purely dynamic axis. Together, these three axes define a coherent biological hierarchy: response geometry is organised primarily by baseline architectural complexity, secondarily by tumour burden, and thirdly by treatment-induced structural change. PC1 was therefore interpreted as a structural-disorder axis rather than a tumour-burden axis: it aligned strongly with baseline structural entropy, the same feature defined in §2.3 (ρ = −0.933), whereas FTV primarily defined the orthogonal tumour-burden dimension on PC2. The resulting manifold structure is shown in Fig. 1; full PC–feature correlation table and extended projections are in Supplementary Table S2.4 and Figs. S2–S3.

Sensitivity analyses confirmed PC1 stability under bootstrap resampling, entropy rank-normalisation, and ablation (Supplementary Section S2.6). As a direct burden-independence test, baseline entropy was residualized against baseline FTV prior to PCA reconstruction; the dominant and secondary axes were preserved at near-identity (PC1 axis r = 0.992; PC2 axis r = 0.988), and the entropy alignment of the dominant axis strengthened rather than attenuated after burden removal (|ρ| = 0.957), confirming that manifold organisation is not a disguised baseline-size effect. Full stability metrics are in Supplementary Section S10.

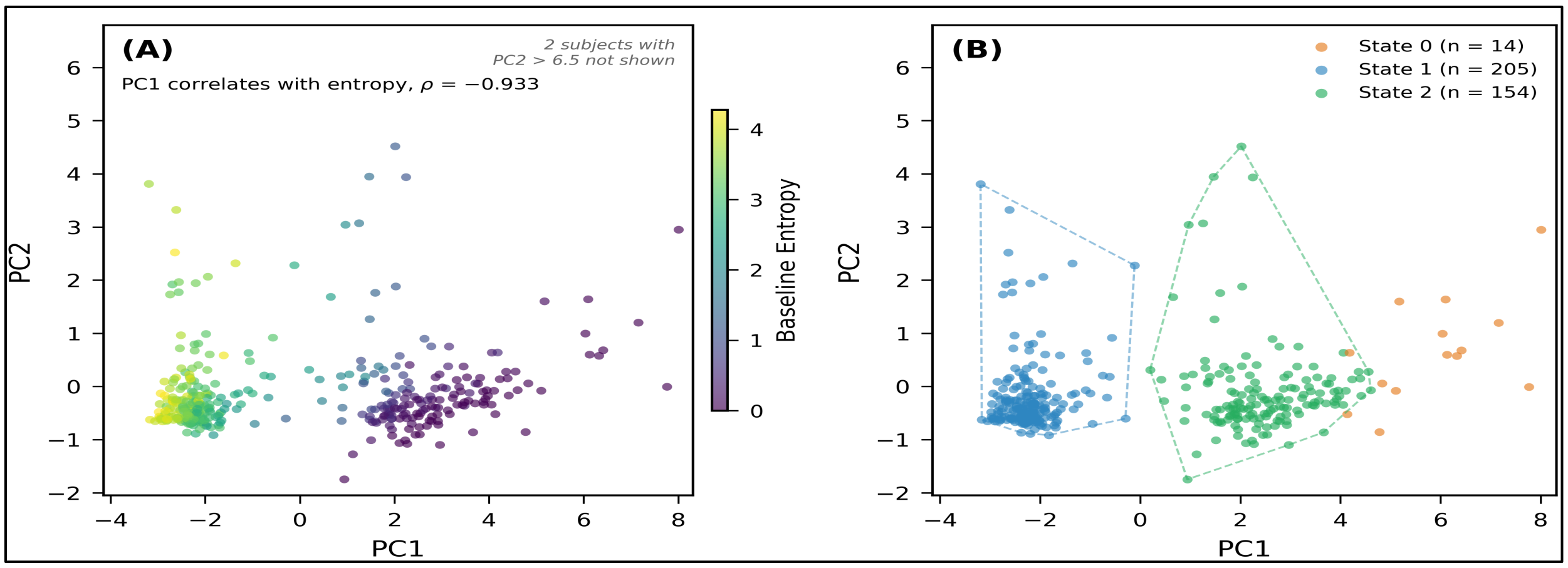

**Fig. 1.** *Unsupervised LDRS manifold structure in the I-SPY2 training partition (n = 373). (A) Principal component projection (PC1 vs. PC2) with subjects coloured by baseline Shannon entropy. PC1 exhibits the strongest single-feature correlation in the 24-dimensional longitudinal vector (Spearman ρ = −0.933), indicating baseline structural heterogeneity as the dominant axis of response space without supervised label guidance. Two subjects with PC2 > 6.5 are excluded from the display range (noted in panel). (B) Three LDRS states identified by k-means clustering (k = 3; bootstrap ARI = 0.980, 200 resamples). Convex hulls are shown for State 1 (n = 205; high entropy, compact trajectories) and State 2 (n = 154; low entropy, heterogeneous volume dynamics). State 0 (n = 14) represents an outlier subgroup with extreme baseline volume. State boundaries emerge from unsupervised geometry; no outcome information was used in manifold construction or clustering.*

#### 3.2.2. Cluster Biological Validation

Three LDRS states were identified with strong internal stability (Supplementary Table S2.1). State separation was overwhelmingly driven by baseline structural heterogeneity rather than tumour burden: entropy variance explained by state membership was $\varepsilon^2 = 0.745$ (KW $H = 277.69$, $p = 5.01\times10^{-61}$; all pairwise comparisons significant after Bonferroni correction), whereas the FTV effect size was negligible ($\varepsilon^2 = 0.018$; Supplementary Table S2.3). State trajectory profiles are shown in Fig. 2; LDRS state biological characteristics are in Table 3. Entropy-defined phenotype groups did not differ meaningfully in age, molecular subtype distribution, or pCR rate (all $p > 0.35$; Supplementary Table S10.3), confirming the entropy phenotype is not a surrogate for standard clinicopathologic grouping.

Cross-tabulation of the pre-specified entropy-defined phenotype against unsupervised LDRS state assignment revealed exceptionally strong concordance ($\chi^2 = 415.3$, $p = 6.73\times10^{-91}$): 88.4% of State 1 subjects were Entropy_high, whereas States 0 and 2 were composed entirely of Entropy_low subjects. This convergence between a pre-specified threshold applied to a single baseline feature and an unsupervised 24-dimensional manifold partition demonstrates that both approaches recover the same dominant biological separation. The manifold additionally resolves structural heterogeneity within the low-entropy regime — distinguishing State 0 (early volumetric instability) from State 2 (burden-dominant progressive phenotype) — a distinction the binary threshold cannot capture.

**Table 3. LDRS State Biological Characteristics (Training Partition, n = 373)**

| Characteristic | State 0 (n = 14) | State 1 (n = 205) | State 2 (n = 154) | KW p | $\varepsilon^2$ |
|---|---|---|---|---|---|
| Median baseline entropy | 0.010 | 3.263 | 0.254 | $5.01\times10^{-61}$ | 0.745 |
| Median baseline FTV, cc | 13.06 | 5.46 | 3.55 | 0.014 | 0.018 |
| Phenotype | Early instability | High structural complexity | Burden-dominant progressive | — | — |

*Full IQR values and Dunn's pairwise test results in Supplementary Table S2.3.*

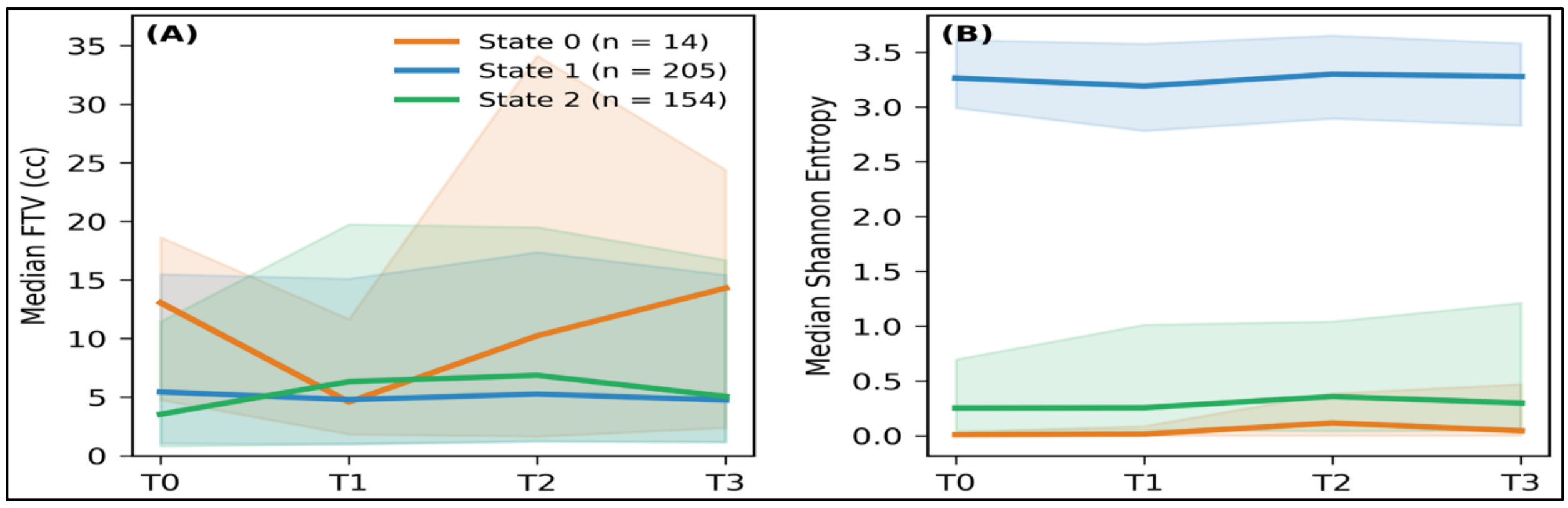

**Fig. 2**. *Longitudinal trajectory profiles by LDRS state in the I-SPY2 training partition (n = 373)***.** *(A) Median functional tumour volume (FTV) across treatment timepoints (T0–T3) with interquartile ranges. State 1 (n = 205) maintains stable low volume throughout treatment. State 2 (n = 154) shows progressive volume reduction. State 0 (n = 14) exhibits high baseline volume with variable trajectory. (B) Median Shannon entropy trajectories with interquartile range bands. State 1 maintains persistently high structural entropy (median > 3.0 at all timepoints), whereas State 2 maintains low entropy throughout, confirming that the manifold separates fundamentally different tissue architectures rather than simple volume categories. The dissociation between volume and entropy dynamics — State 1 has low volume but high entropy, State 2 has higher volume but low entropy — suggests that structural heterogeneity and tumour burden capture complementary biological dimensions of treatment response.*

### 3.2.3. DLRM Validation: Biological Organisation Is Intrinsic to the Data

Whether the biological interpretation of manifold axes reflects structure intrinsic to the data or merely post-hoc labelling of generic variance components was tested formally by comparing a biologically constrained manifold (DLRM) against the original unconstrained PCA solution. DLRM explicitly imposed the entropy/burden/remodelling axes as construction constraints using sequential biologically anchored partial least squares, with no outcome labels at any stage. This test addresses a gap present throughout the unsupervised radiomic manifold literature, where axis interpretation is routinely assigned post hoc without formal verification of intrinsicness.

The constrained and unconstrained solutions converged at near-identity: score correlations between DLRM and PCA axes were |r| = 0.996 (structural heterogeneity), |r| = 0.997 (tumour burden), and |r| = 0.952 (remodelling) on the training partition, with consistent held-out test values; downstream state assignments were identical under both methods (ARI = 1.000; full statistics in Supplementary Table S2.9). The near-identity of constrained and unconstrained solutions establishes that the entropy-dominated axis organisation is intrinsic to the longitudinal response geometry: had the biological interpretation been incidental to PCA's variance maximisation, imposing it as a construction constraint would have produced a different manifold.

### 3.3. Orthogonality to Molecular and Clinical Prediction Structure

#### 3.3.1. Molecular Independence

The entropy-defined structural phenotype showed no meaningful association with any of four independent molecular and clinical classifiers. Against receptor-defined subtype, LDRS state membership showed no significant association ($\chi^2(6) = 7.55$, $p = 0.273$, Cramér's V = 0.10). Against PAM50 intrinsic subtype from matched pretreatment gene expression in 704 subjects (GSE194040 [58]), entropy group showed no meaningful association ($\chi^2 = 4.87$, $p = 0.301$, Cramér's V = 0.083), with no correlation between continuous baseline entropy and proximity to any PAM50 centroid (all Spearman $\rho < 0.06$, all $p > 0.13$; Supplementary Section S2.7.2). Against MammaPrint genomic risk [50], entropy group showed no meaningful association (Cramér's V = 0.072). Against baseline tumour burden, FTV-adjusted manifold reconstruction confirmed independence (PC1 axis r = 0.992 after direct FTV removal; §3.2.1).

Within every stratum defined by each classifier, entropy separation remained highly significant: within all four HR/HER2-defined subtype groups (all $p < 10^{-7}$), within all four PAM50 subtypes, and within both MammaPrint strata (Supplementary Fig. S5; Tables S2.8a–c and Section S2.7).

To quantify the irreducibility of this structural dimension, we evaluated whether the dominant manifold axis (PC1) could be recovered from the full conventional baseline stack. A linear model including tumour burden (FTV_T0), age, receptor subtype, MammaPrint, treatment arm, and PAM50 intrinsic subtype explained essentially none of PC1 out of sample (cross-validated $R^2 = -0.023$, indicating the conventional stack performs worse than a constant-mean predictor for PC1 recovery). After addition of baseline entropy alone, the same model achieved cross-validated $R^2 = 0.876$ ($\Delta$CV $R^2 = 0.899$; Fig. 3). The dominant manifold axis is therefore not a disguised reformulation of known clinical or genomic categories. It is a distinct structural dimension that the entire conventional baseline stack — including both receptor-defined and transcriptionally derived molecular classifiers — cannot access. Although PC1 aligned strongly with baseline entropy (Spearman $\rho = -0.933$), the full conventional stack left it essentially unexplained, confirming that the latent structural phenotype is orthogonally invisible to the established classification architecture (Supplementary Table S13.2).

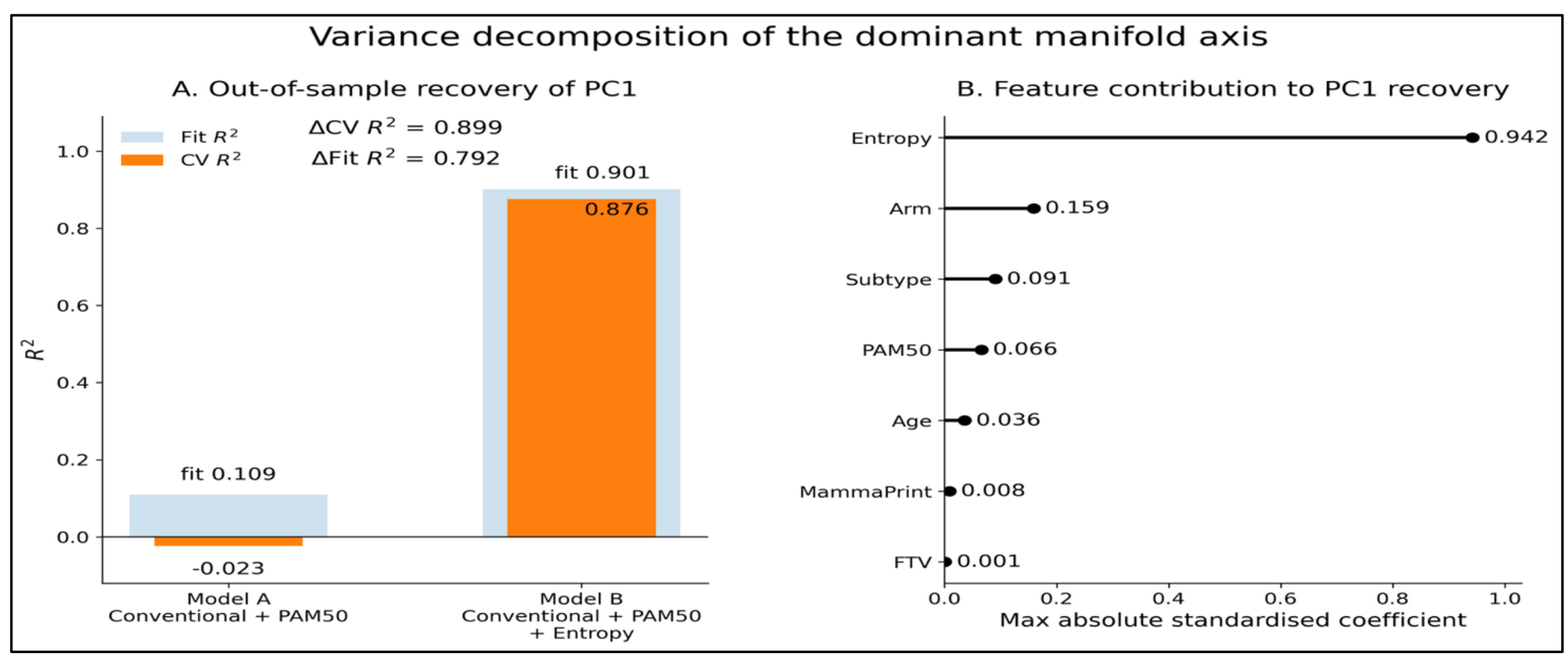

**Fig. 3.** *The conventional clinical and genomic stack cannot recover the dominant manifold axis; baseline entropy can. (A) Cross-validated $R^2$ for Model A (full conventional baseline stack: FTV, age, receptor subtype, MammaPrint, treatment arm, PAM50) versus Model B (conventional stack + baseline entropy). Model A explains negative variance out of sample (CV $R^2$ = -0.023); adding entropy alone raises recovery to CV $R^2$ = 0.876 (Delta = 0.899). (B) Feature-level contribution showing that baseline entropy dominates PC1 recovery while all conventional descriptors contribute negligibly. The dominant manifold axis contains structural information that the entire clinical and genomic classification stack cannot access.*

### 3.3.2. Subtype-Adjusted pCR Signal Test

A pre-specified signal test evaluated whether the pre-specified manifold contributed meaningful pCR discrimination beyond an explicit molecular baseline (HR + HER2 + MammaPrint) in the trajectory-complete cohort (training n = 373; test n = 166).

The molecular baseline achieved a held-out test AUROC of 0.709. Adding the manifold's PC1, baseline entropy, or LDRS state produced no statistically significant improvement: no addition reached significance (Table 4). The manifold is not primarily a subtype-conditioned pCR augmenter. Its principal clinical relevance lies in structural response geometry and recurrence-risk structuring — a framing directly motivated by the pCR–survival dissociation examined in §3.3.4 and §4.3. Full model outputs and coefficient stability are provided in Supplementary Section S5.11.

**Table 4. Subtype-Adjusted pCR Signal Test:Held-Out Test Performance (Trajectory-Complete Cohort,n= 166)**

| Model | Test AUROC | ΔAUROC vs Base | Bootstrap 95% CI | DeLong p | Test AUPRC |
|---|---|---|---|---|---|
| Base: HR + HER2 + MP | 0.709 | — | — | — | 0.522 |
| Base + PC1 | 0.714 | +0.005 | [−0.028, +0.038] | 0.757 | 0.614 |
| Base + Entropy_T0 | 0.709 | +0.001 | [−0.027, +0.028] | 0.970 | 0.590 |
| Base + LDRS state | 0.698 | −0.010 | [−0.036, +0.016] | 0.424 | 0.544 |
| Base + PC1 + LDRS | 0.687 | −0.021 | [−0.058, +0.014] | 0.240 | 0.552 |

*All models trained on trajectory-complete training partition (n = 373, pCR rate 0.308) and evaluated once on the locked held-out test partition (n = 166, pCR rate 0.307). Bootstrap CIs from 2,000 stratified resamples of test predictions. DeLong p values are paired comparisons against the molecular baseline. MP = MammaPrint.*

### 3.3.3. Entropy-Defined Phenotype Organises Longitudinal Response Trajectories

Entropy-defined separation was preserved throughout the entire treatment course between Entropy_high (n = 343) and Entropy_low (n = 363) groups (all $p < 10^{-59}$ at T0–T3), while functional tumour volume differences — present at baseline (p = 0.002) — were completely absent by T1 and remained absent through T3 (p = 0.93). This dissociation, persistent structural separation with vanishing volumetric separation, demonstrates that entropy captures a dimension of tumour organisation not reducible to burden dynamics under therapy.

Entropy_high tumours showed early volumetric reduction (median ΔFTV_10 = −0.733 versus +0.208 in Entropy_low; $p = 5.76\times10^{-4}$) and consistent structural homogenization across all treatment intervals (all $p \leq 3.96\times10^{-6}$), while Entropy_low tumours showed modest entropy increases — a divergence consistent with distinct patterns of treatment-induced tissue remodelling. Within-subtype analysis confirmed longitudinal separation across all four molecular subtype groups at all timepoints (all $p < 10^{-5}$; Supplementary Fig. S17 and Section S11.1), establishing that the entropy axis organises response geometry within receptor-class biology rather than merely recapitulating between-subtype partitioning. Longitudinal trajectory profiles are shown in Fig. 4.

Beyond longitudinal separation between groups, the structural phenotype was directly persistent as an individual-level property. Continuous baseline-to-endpoint entropy rank correlation was strong (Spearman $\rho = 0.736$, $p < 10^{-64}$; Pearson r = 0.789), while 88.2% of subjects retained their median-split structural state through treatment (329/373; transition matrix: 164 low-to-low, 165 high-to-high, 22 crossovers in each direction). Ordinal persistence was also strong: 93.6% of subjects remained in the same or adjacent entropy quartile from baseline to post-treatment assessment. The near-zero mean signed entropy change (0.028) with substantial absolute change (0.633) indicates that treatment perturbs individual structural values but does not systematically reorganise the population's structural hierarchy. The entropy-defined phenotype is not transiently reconfigured by therapy; instead, baseline structural ordering is strongly preserved through the full treatment course, supporting the interpretation that the manifold captures an enduring biological property rather than a temporary response-state fluctuation (Supplementary Table S13.3).

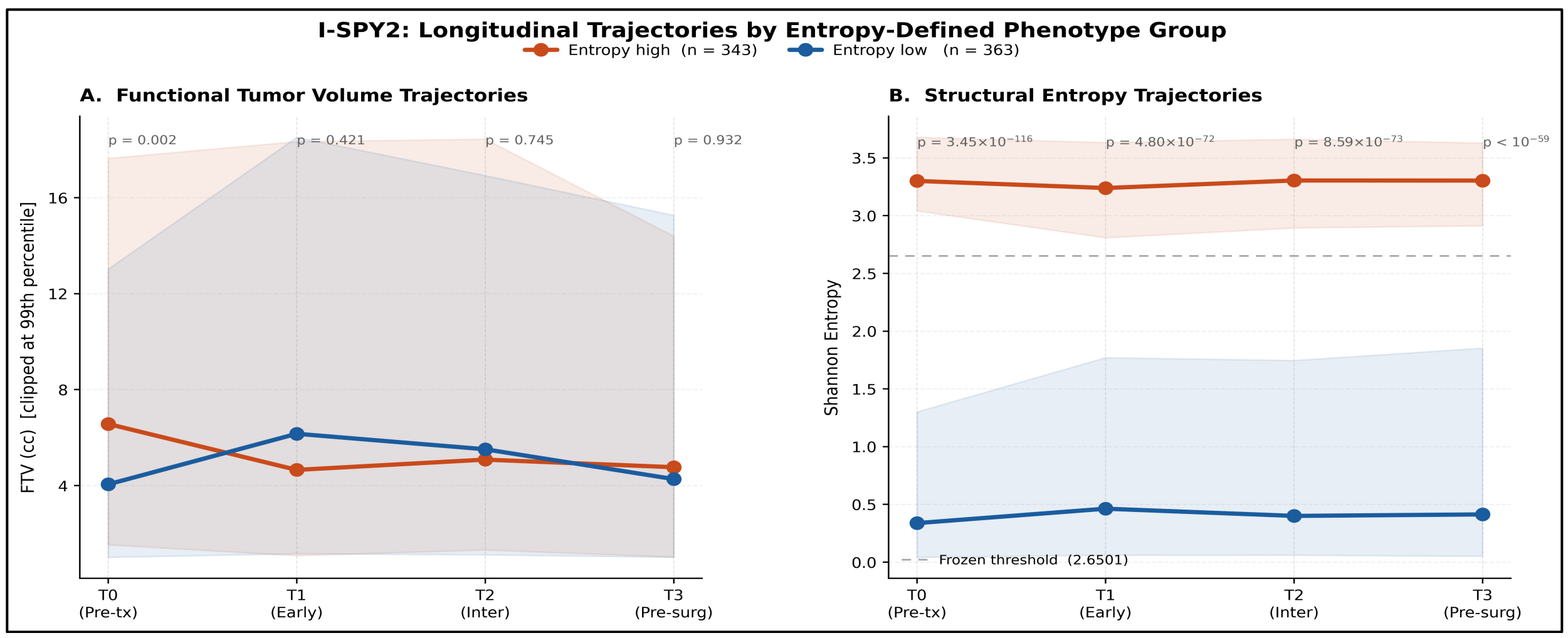

**Fig. 4.** *Longitudinal functional tumour volume and structural entropy trajectories by entropy-defined phenotype group in the I-SPY2 longitudinal analysis cohort (n = 706). Entropy_high (n = 343, orange) and Entropy_low (n = 363, blue) groups are defined by the pre-specified I-SPY2 training-partition median threshold (2.6500518). Median trajectories are shown with interquartile range bands across four treatment timepoints: T0 (pre-treatment), T1 (early), T2 (inter-regimen), and T3 (pre-surgery). Mann–Whitney U p-values are annotated at each timepoint. (A) Functional tumour volume differences between groups, present at baseline (p = 0.002), were absent by T1 and remained absent through T3 (p = 0.421, 0.745, 0.932). (B) Shannon entropy separation was highly significant at all four treatment timepoints (p = $3.45\times10^{-116}$, $4.80\times10^{-72}$, $8.59\times10^{-73}$, $1.13\times10^{-59}$ at T0–T3). The dashed horizontal line indicates the pre-specified threshold (2.6500518). This dissociation — persistent structural separation with vanishing volumetric separation — establishes entropy as a dynamic organiser of response geometry rather than a static baseline descriptor.*

### 3.3.4. Entropy Phenotype Organises Structural Trajectories Within pCR Achievers

Entropy-defined groups remained strongly separated in longitudinal structural entropy across all four treatment timepoints within the pCR-positive subgroup (n = 220; Entropy_high n = 107, Entropy_low n = 113) despite uniformly favourable pathologic response (all $p < 10^{-19}$ at T0–T3). Functional tumour volume showed no meaningful separation at any timepoint (all p > 0.069). Within-pCR trajectory plots are provided in Supplementary Fig. S18.

This dissociation — persistent structural separation with absent volumetric separation among patients with identical surgical outcomes — provides direct imaging-visible evidence of the mechanism by which structural response biology escapes pCR resolution. Critically, within the pCR population, higher baseline entropy was associated with more favourable early volumetric response rather than volumetric resistance: baseline entropy correlated inversely with early percent volume change (Spearman ρ = -0.265, p = $8.0 \times 10^{-5}$, n = 217), and the highest-entropy quartile showed substantially stronger early shrinkage than the lowest-entropy quartile (percent-change p = $1.66 \times 10^{-4}$; raw ΔFTV_10 p = 0.0029). This association persisted after adjustment for baseline tumour burden (size-adjusted ρ = −0.262, p = $9.2 \times 10^{-5}$), confirming that the pattern is not merely a larger-tumours-shrink-more artefact. The most favourable early volumetric responses were enriched for persistent structural disorder, not depleted of it. Apparent volumetric success can therefore coexist with — and obscure — that disorder within pCR, identifying a specific mechanism by which standard response language can overstate biological completeness (Supplementary Table S13.4). If this structural divergence carries recurrence-relevant consequences, it identifies a candidate basis for differentiating post-surgical surveillance intensity within the pCR-positive population — the subgroup in which existing endpoint-based stratification provides the least biological resolution. This implication requires prospective validation; the Duke survival finding is consistent with this interpretation and is examined in section 4.3.

### 3.4. Incremental Prediction Ladder

An incremental evidence ladder (M0-M5) evaluated under a locked train/test split confirmed that molecular subtype markers provide the dominant predictive signal (M0 to M1: ΔAUROC = +0.147) and that baseline imaging features contribute significant incremental discrimination (M1 to M2: ΔAUROC = +0.059). Later additions — LDRS state (M4) and full-course imaging (M5) — produced smaller non-significant gains, culminating in a ceiling of AUROC = 0.781, consistent with baseline tumour biology capturing the dominant signal and likely sample-size-limited rather than model-class-limited. Predicted probabilities were well calibrated (ECE less than 0.10). The full ladder, model definitions, and performance metrics are in Supplementary Table S13.7 and Supplementary Figure S13.1. The subtype-adjusted signal test (section 3.3.2) confirmed that the manifold does not materially augment pCR discrimination once molecular subtype is known; its primary clinical expression is recurrence-risk organisation rather than endpoint reclassification.

### 3.5. External Validation of the Longitudinal Response Manifold

Of 64 UCSF subjects identified, 49 had complete T0–T2 FTV trajectories eligible for survival analysis (15 excluded for incomplete trajectories prior to outcome analysis; Supplementary Section S6.1). Baseline FTV differed approximately four-fold from I-SPY2 training (20.7 cc versus 5.0 cc; Cohen's d = −0.35; KS $p = 5.0\times10^{-8}$), with separation narrowing progressively during therapy (Supplementary Table S6.2; Fig. S9). The external manifold was evaluated under documented distributional shift rather than in a cohort selected for imaging similarity.

Projection of 49 UCSF subjects through locked transforms yielded State 0 (n = 7), State 1 (n = 40), and State 2 (n = 2). The primary comparison used the pre-specified binary grouping of State 1 versus Non-State 1 (n = 9). An ordered risk gradient was observed across all three projected states (State 1: 25% event rate; State 0: 43%; State 2: 100%) despite states being defined without survival labels in an entirely separate institution. The primary survival analysis demonstrated statistically significant separation (log-rank $\chi^2 = 8.20$, $p = 0.0042$; absolute recurrence rate difference 31 percentage points; Table 5; Fig. 5), achieved under strict projection-only inference with no retraining, recalibration, or parameter adjustment on UCSF data. Adjusted Cox regression was underpowered at 15 events and a 40/9 group allocation (achieved power = 13.6% for the observed hazard ratio), yielding a directionally consistent but wide estimate (HR = 1.44, 95% CI 0.76–2.73; Supplementary Section S6.4); the primary inferential basis is therefore the convergent robustness evidence (log-rank $p = 0.0042$; RMST advantage 22.6 months, permutation $p = 0.002$; leave-one-out all 49 iterations $p < 0.05$) rather than the adjusted Cox estimate alone.

Robustness checks (bootstrap resampling, permutation testing, and leave-one-out influence analysis) confirmed that the survival signal is distributed across the cohort and robust to the removal of any individual or event subject; full statistics are in Supplementary Section S6.5. RMST analysis [62] at a 60-month horizon yielded a State 1 advantage of 22.6 months (bootstrap 95% CI 5.7–39.5 months; permutation $p = 0.002$), confirming survival separation on an absolute scale without proportional-hazards assumptions.

Pre-specified geometry and confounding audits established that UCSF state assignments were geometrically supported and biologically coherent: 91.8% of subjects projected into the supported manifold region; Non-State 1 subjects occupied more peripheral positions (median centroid distance 1.399 versus 0.474 for State 1; $p = 0.0024$); projected states recovered markedly different longitudinal FTV trajectories within UCSF (Non-State 1 approximately six-fold higher baseline FTV; $p = 4\times10^{-6}$), confirming that the pre-specified projection captured real trajectory structure. Molecular confounding analyses showed no HR, HER2, or subtype imbalance across states (all $p > 0.10$), and pCR rates were essentially identical (12.5% versus 11.1%; Fisher $p = 1.000$), ruling out pCR as an explanatory mechanism. Within UCSF State 1, baseline entropy differed significantly between eventual recurrers and non-recurrers ($p = 0.035$), indicating recurrence-relevant structural heterogeneity persists within a manifold-defined state. Full audit statistics are in Supplementary Sections S6.9–S6.13.

**Table 5. Primary External Validation Survival Analysis (UCSF, n = 49)**

| Group | n | RFS Events | Event Rate | Log-rank $\chi^2$ | p-value | Absolute Risk Difference |
|---|---|---|---|---|---|---|
| State 1 | 40 | 10 | 25% | | | |
| Non-State 1 | 9 | 5 | 56% | | | |
| Combined | 49 | 15 | — | 8.20 | 0.0042 | 31 percentage points |

*Log-rank $\chi^2$, p-value, and absolute risk difference refer to the State 1 vs Non-State 1 comparison.*

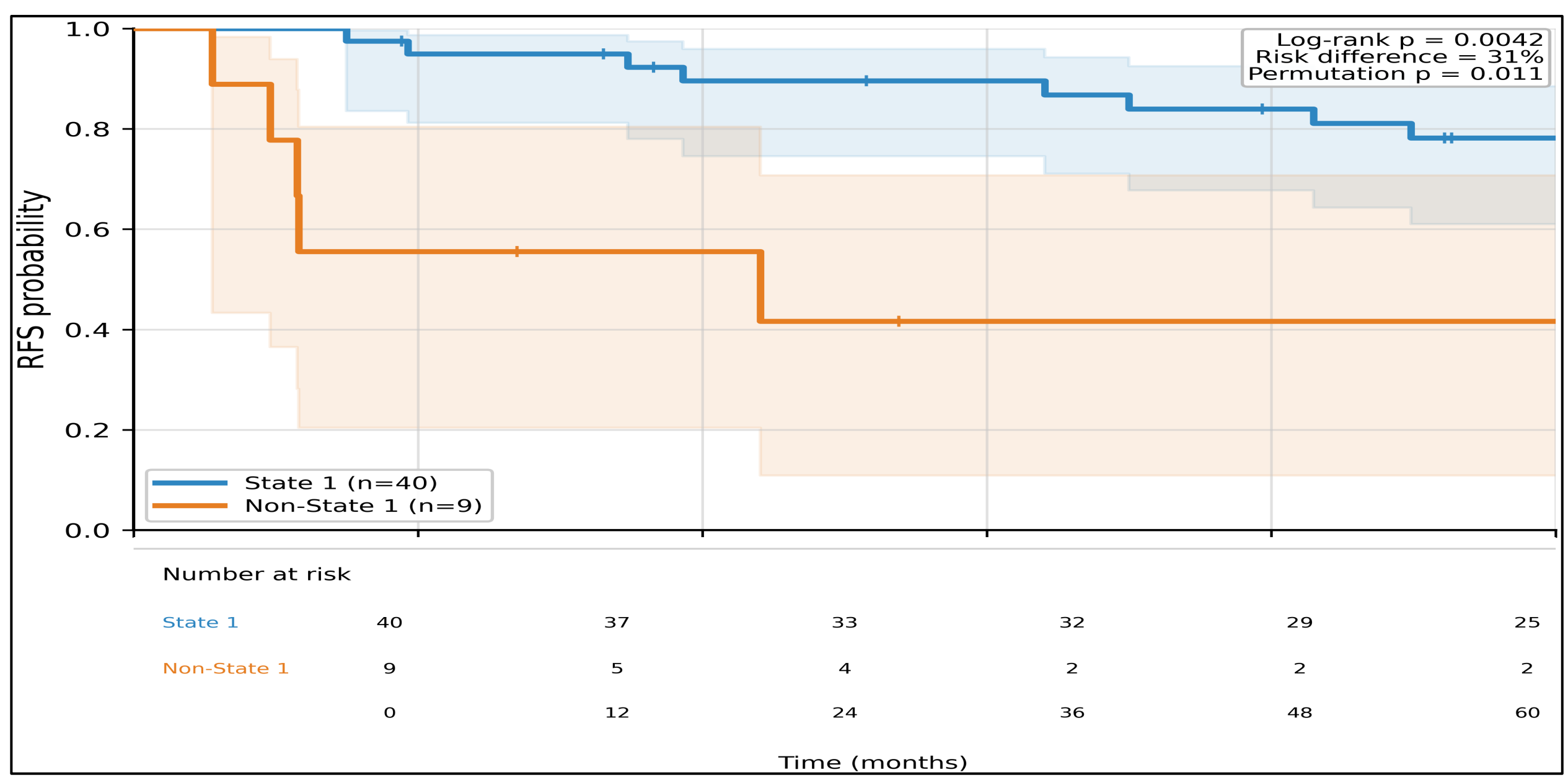

**Fig. 5.** *External prognostic validation: recurrence-free survival by locked LDRS state assignment in the independent UCSF cohort (n = 49; 1995–2002, GE 1.5T). Subjects were projected into the pre-specified Tier-2 reduced manifold (FTV-only, three timepoints) using pre-specified I-SPY2 training transforms without parameter re-estimation or model recalibration. State 1 (n = 40) demonstrated significantly higher recurrence-free survival compared with Non-State 1 (n = 9; log-rank p = 0.0042; absolute five-year risk difference = 31 percentage points). Permutation testing (1,000 resamples) confirmed that the observed separation exceeds chance (permutation p = 0.011). Number-at-risk table is shown below the curves.*

### 3.6. I-SPY1 External Tier-2 Validation: Feature-Family Specificity and Locked Geometry Transport

The I-SPY1 (ACRIN 6657) external arm addressed a methodological question distinct from UCSF and Duke — whether feature-family specificity of external transport is demonstrable within a single cohort. Tumour-volume extraction from archived DICOM SEG objects produced a 126-subject T0/T1/T2 cohort with 39 recurrence events (Supplementary Section S9.1).

**Entropy fails commensurability while FTV passes, in one cohort.** In the same I-SPY1 cohort where entropy extraction failed RFCA Stage 1 (Cohen's d = 1.354 after forensic cleanup; Supplementary §S8.14), extracted FTV dynamics occupied a practically commensurable Tier-2 regime (ΔFTV_10 d = −0.039; ΔFTV_20 d = 0.030). In the same cohort, under the same configuration: entropy failed commensurability screening while FTV dynamics passed — proving that transportability is a representation-family property, not a cohort-level property.

**Locked geometry transport and state recovery.** Locked projection succeeded for all 126 subjects, with 87.3% projecting into the 95th-percentile support region of the training distribution — comparable to UCSF (91.8%). Three clinically coherent response phenotypes were recovered: high-burden shrinking (State 0, n = 26), low-burden modest-response (State 1, n = 90), and progression-like (State 2, n = 10). State 1 was fully supported within the training manifold (100% within 95th-percentile support; median centroid distance 0.541); States 0 and 2 occupied substantially more peripheral positions (median distances 3.023 and 2.785 respectively).

**pCR-trajectory dissociation replicated.** Despite marked between-state trajectory divergence (Kruskal-Wallis p ≈ 0 for all FTV variables), pCR rates were flat across states (29.2%, 26.7%, 30.0%; $\chi^2$ p = 0.956), while HR-status showed meaningful enrichment across states (p = 0.026). This replicates the paper's central I-SPY2 finding across a structurally independent acquisition-era cohort.

**Coarse survival and within-state microstructure.** Coarse binary survival separation was not reproduced (log-rank p = 0.801), mechanistically explained by the peripheral manifold placement of Non-State 1 subjects rather than absence of biological signal. Within the fully in-support State 1 (n = 90, 29 recurrence events), centroid distance independently predicted recurrence-free survival (HR per unit = 2.04, 95% CI 1.15–3.64, p = 0.015), as did baseline FTV (HR per 1 cc = 1.019, p = 0.016), establishing that manifold placement geometry encodes recurrence-relevant microstructure within the best-supported state. Full I-SPY1 audit statistics are in Supplementary Section S9.

To test whether the manifold carried graded clinical meaning beyond discrete state assignments, we evaluated continuous projection coordinates against recurrence-free survival under explicit support conditioning. In the full I-SPY1 projected cohort (n = 126, 39 events), the overall continuous PC1 association was directionally consistent but not significant (HR per 1 SD = 0.805, 95% CI 0.606-1.070, p = 0.135). However, after restriction to the RFCA-supported region (n = 110, 36 events, Within95Support = TRUE), continuous PC1 became significantly associated with recurrence-free survival (HR per 1 SD = 0.624, 95% CI 0.453-0.861, p = 0.004; concordance index 0.628; Fig. 6). Within the same supported subset, assigned centroid distance also retained prognostic value (HR = 1.404, 95% CI 1.033-1.909, p = 0.030), indicating that both axial position and geometric deviation within the manifold carry outcome-relevant graded information. Out-of-support subjects occupied a markedly different geometric regime (Cohen d up to 3.65 across distance-based features), demonstrating that the support split reflects a substantive manifold boundary rather than a post-hoc analytic convenience. In the in-support HER2-positive subgroup (n = 32, 14 events), geometric risk remained evident (centroid distance HR = 1.733, 95% CI 1.040-2.889, p = 0.035). This result demonstrates that RFCA support is not merely an audit criterion but a prospective predictor of where continuous manifold geometry carries clinical meaning (Supplementary Table S13.5). External validation is therefore not binary; it is representation-conditioned, and the RFCA framework provides the pre-specified logic for determining where transfer-based inference remains interpretable.

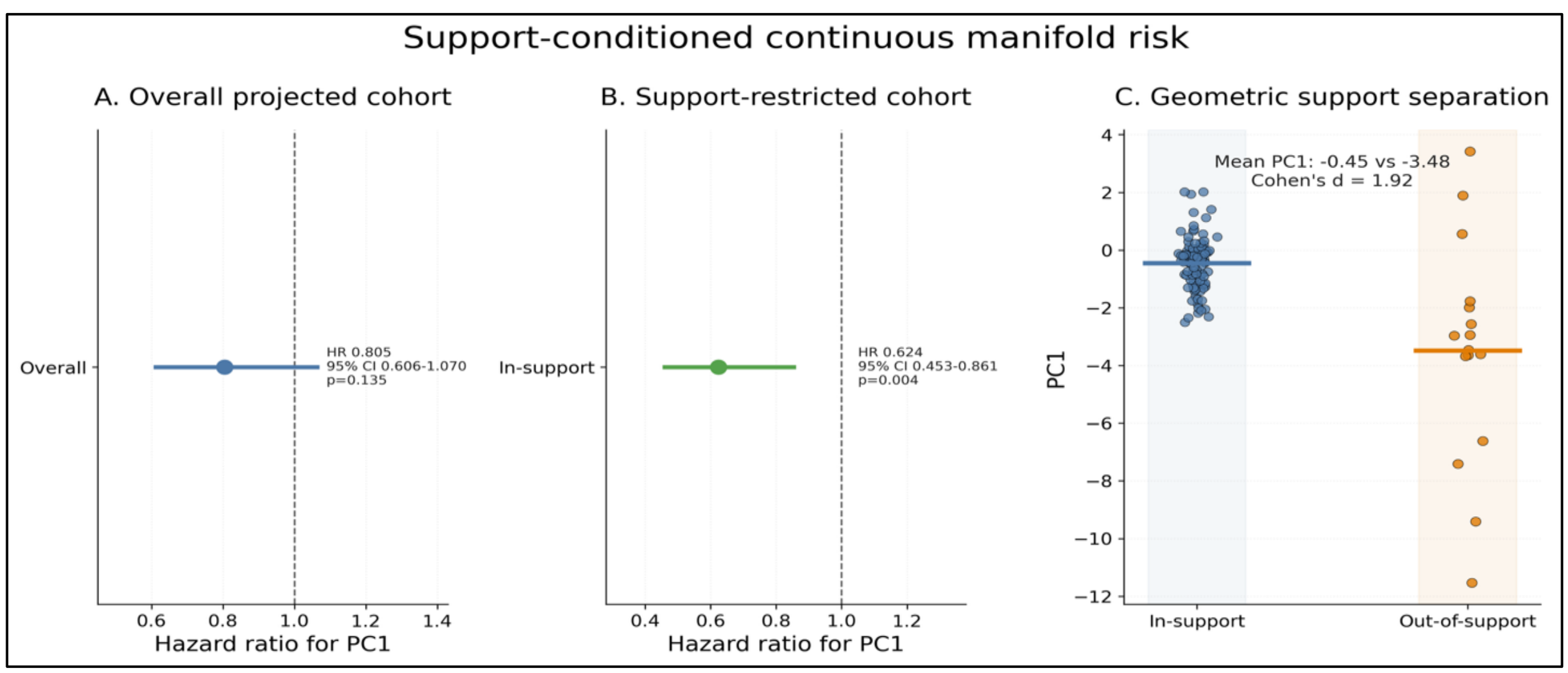

**Fig. 6.** *Support-conditioned continuous manifold risk in I-SPY1 (n = 126). (A) Overall continuous PC1 vs recurrence-free survival: HR = 0.805, p = 0.135 (non-significant). (B) In-support continuous PC1 vs recurrence-free survival (n = 110, Within95Support = TRUE): HR = 0.624, p = 0.004 (significant). The transition from non-significant overall to significant within RFCA-supported regions demonstrates that the four-stage commensurability framework correctly predicts where manifold geometry carries clinical meaning. (C) In-support vs out-of-support geometric distribution showing markedly different projection regimes (Cohen d up to 3.65).*

### 3.7. Duke Cohort Characteristics

The DRFS-eligible Duke cohort comprised 908 subjects with 76 distant recurrence events across four molecular subtype groups (HR+/HER2− n = 587, 39 events; TNBC n = 162, 23 events; HR+/HER2+ n = 101, 8 events; HR−/HER2+ n = 58, 6 events). Application of the pre-specified I-SPY2 training-partition entropy threshold (2.6500518) classified 709 subjects as Entropy_low and 199 as Entropy_high (Supplementary Tables S8.1–S8.2). A Duke-internal median split (2.0226), used as a threshold sensitivity analysis, produced balanced groups of 454 each (Entropy_low 51 events, 11.2%; Entropy_high 25 events, 5.5%).

### 3.8. Duke Transportability Audit

#### 3.8.1. BHI Composite Commensurability

Systematic feature-label matching of all eleven I-SPY2 BHI components against the 530-column Duke feature matrix identified only one matched feature (9.1% match rate; Supplementary Table S8.3), rendering full composite projection invalid under the pre-specified decision rule (§2.10). The RFCA was therefore applied to the four individual Duke entropy-family representations sharing the label class of the BHI's structural anchor (Entropy_T0).

#### 3.8.2. Four-Entropy Representation-Family Matrix

Table 6 and Fig. 7 present the audit across all four Duke entropy families against both reference distributions; complete statistics and bootstrap rank-stability results are in Supplementary Tables S8.4–S8.5. Against the I-SPY2 training reference, Entropy_Tumor was the sole REGIME_COMPATIBLE representation (Cohen's d = −0.020; P(rank 1) = 1.000). Against the UCSF reference, the hierarchy inverted completely: PE_map_Entropy_tumor became rank 1 (d = 0.451) while Entropy_Tumor fell to rank 4 (d = 6.634), with this inversion deterministic across all 2,000 bootstrap resamples (P(rank 1) = 1.000 for both rank-1 families). The transportability unit is therefore the representation family, not the feature label: entropy computed from raw tumour intensity and entropy computed from percent-enhancement maps are not interchangeable constructs, and their external validity depends on the reference regime. Cross-software calibration confirmed central alignment (mean z-score 0.017) with approximately 50% dynamic-range compression in Duke (z-score SD 0.494 versus expected 1.0).

**Table 6. Cross-Cohort Representation-Family Transportability Audit**

| Duke Entropy Family | Duke Mean | vs I-SPY2 Train d | Verdict | Rank | vs UCSF d | Verdict | Rank |
|---|---|---|---|---|---|---|---|
| Entropy_Tumor | 2.034 | −0.020 | REGIME_COMPATIBLE | 1 | 6.634 | NOT_COMPATIBLE | 4 |
| SER_map_Entropy_tumor | 4.343 | −0.908 | PARTIALLY_SHIFTED | 2 | 0.924 | PARTIALLY_SHIFTED | 2 |
| PE_map_Entropy_tumor | 5.918 | −1.847 | PARTIALLY_SHIFTED | 3 | 0.451 | REGIME_COMPATIBLE | 1 |
| WashinRate_map_Entropy_tumor | 8.755 | −7.482 | NOT_COMPATIBLE | 4 | −3.780 | NOT_COMPATIBLE | 3 |

*I-SPY2 training reference: n = 373, mean = 2.013, SD = 1.506. UCSF reference: n = 58, mean = 6.933. This n = 58 denotes UCSF T0 subjects with non-missing entropy used for the representation-family comparison, distinct from the n = 49 complete-trajectory cohort used for projection and survival. Bootstrap rank-stability: P(rank 1) = 1.000 for both rank-1 families across 2,000 stratified resamples. Verdict thresholds: REGIME_COMPATIBLE |d| < 0.50; PARTIALLY_SHIFTED 0.50 ≤ |d| < 2.00; NOT_COMPATIBLE |d| ≥ 2.00.*

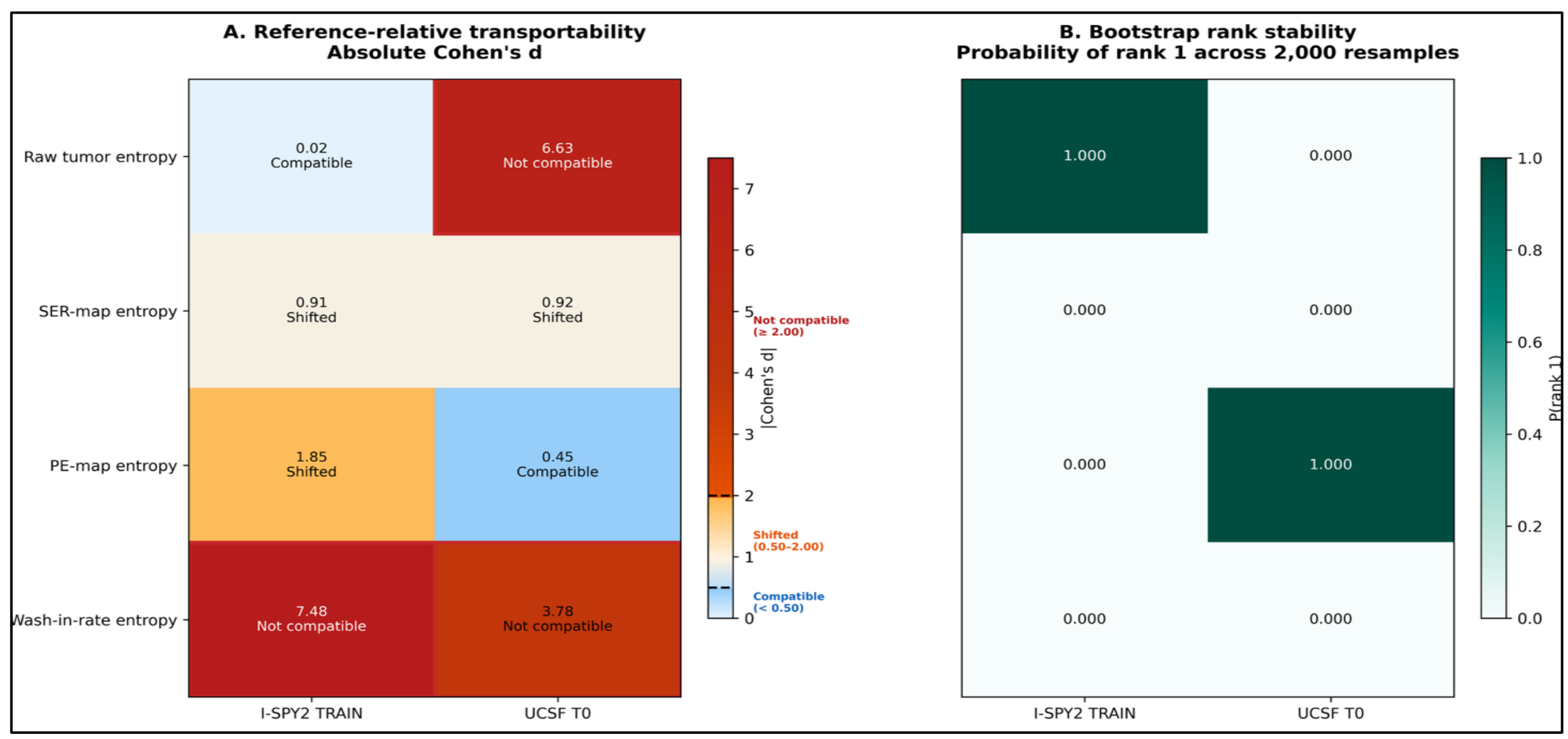

**Fig. 7.** *Representation-family-dependent, reference-regime-dependent transportability of Duke entropy features. (A) Absolute Cohen's d heatmap comparing four Duke entropy representation families against the I-SPY2 TRAIN and UCSF T0 reference entropy distributions. Lower values indicate closer numeric commensurability; dashed lines on the colorbar mark the RFCA Stage 1 verdict boundaries (|d| = 0.50 and |d| = 2.00); red-bordered cells mark families classified as NOT_COMPATIBLE in the locked Duke transportability audit. Relative to I-SPY2 TRAIN, raw tumour entropy was the closest Duke family, followed by SER-map, PE-map, and wash-in-rate entropy. Relative to UCSF T0, the ordering inverted, with PE-map entropy closest and raw tumour entropy most distant. (B) Bootstrap rank-stability heatmap showing the probability that each Duke entropy family ranked closest to each reference cohort across 2,000 stratified bootstrap resamples. The hierarchy was deterministic in both directions: raw tumour entropy ranked first in 100% of resamples relative to I-SPY2 TRAIN, whereas PE-map entropy ranked first in 100% of resamples relative to UCSF T0. Together, these results show that cross-cohort transportability is governed by the pairing between a specific entropy representation family and a specific reference regime, rather than by the nominal feature label "entropy" alone.*

### 3.8.3. MRI Technical Covariate and Grade Independence

Entropy_Tumor showed minimal field-strength dependence (Pearson r = −0.088; Supplementary Table S8.6) and no meaningful association with Nottingham grade (unadjusted Spearman ρ = −0.019, p = 0.633; subtype-adjusted ρ = −0.009, p = 0.812; Supplementary Section S8.12), confirming that the entropy-defined imaging phenotype is not reducible to histopathologic grade.

### 3.8.4. I-SPY1 as Stage 1 Empirical Stress Test

I-SPY1 represents the maximally challenging RFCA Stage 1 scenario: it shares ACRIN acquisition lineage, DCE-MRI infrastructure, feature nomenclature, and PyRadiomics configuration with I-SPY2. Despite this surface similarity, cohort-scale raw-DCE SER recomputation in 38 eligible subjects revealed that the cohort-level entropy distribution remained PARTIALLY_SHIFTED (Cohen’s d = 1.354 after forensic cleanup; Supplementary Section S8.14). All stored-map constructions failed Stage 1 at substantially larger distances (stored SER: d = 4.78 SD; stored PE1: d = 2.57 SD). I-SPY1 was therefore classified as technically feasible but non-commensurate for full entropy-manifold projection. The framework correctly stopped a projection that would have been scientifically invalid despite appearing superficially justified.

### 3.8.5. Stage 4 Extraction-Geometry Coupling Audit (Duke)

Despite passing Stage 1 (Cohen's d = −0.020), Entropy_Tumor showed a strong negative correlation with TumorMajorAxisLength_mm (Spearman ρ = −0.729), indicating that Duke's bounding-box extraction regime introduced substantial spatial-extent coupling absent from the contour-based I-SPY2 extraction. Adding TumorMajorAxisLength_mm as a covariate attenuated the entropy association to non-significance while preserving its protective direction (HR = 0.839, 95% CI 0.401–1.756, p = 0.642), consistent with partial collinearity between structural heterogeneity and spatial extent under bounding-box segmentation rather than confounding that reverses the primary association; notably, this coupling was specific to imaging-derived extent, as Entropy_Tumor showed no association with clinical tumour size (ρ = +0.022, p = 0.559). Under more stringent linear residualization of entropy on major-axis length the association was fully attenuated to the null (HR = 1.11, p = 0.603; degree-2 polynomial HR = 1.08, p = 0.726). The same attenuation pattern extended across all three structural Duke entropy-family representations, establishing that extent entanglement is representation-family level rather than feature-specific. Duke Entropy_Tumor was therefore classified as REGIME_COMPATIBLE but GEOMETRY_ENTANGLED — a diagnostic success: the framework identified a boundary condition that standard validation would have reported as a positive result, establishing that Stage 1 passage alone is insufficient for external biological inference when extraction geometries differ. Full audit statistics are in Supplementary Sections S8.15–S8.19.

### 3.9. Duke External Validation — Distant Recurrence-Free Survival

The Duke cohort serves two analytically linked functions: external DRFS validation of the structural phenotype under a pre-specified, training-derived threshold, and RFCA Stage 4 case study of extraction-geometry entanglement in a bounding-box regime absent from the I-SPY2 contour-based extraction. Both findings are reported below. Among 908 DRFS-eligible Duke subjects (76 distant recurrence events), the pre-specified I-SPY2 training-partition entropy threshold separated distant recurrence-free survival (log-rank p = 0.022; unadjusted HR = 0.452, 95% CI 0.225–0.906, p = 0.025). In the primary adjusted model, restricted to reliably available receptor covariates (Entropy_high with HR and HER2 status; events-per-variable 25.3), the protective association persisted (HR = 0.466, 95% CI 0.232–0.935, p = 0.0317; Supplementary Table S8.8a). A subtype-adjusted specification gave a consistent estimate (HR = 0.463, 95% CI 0.231–0.930, p = 0.031; Supplementary Table S8.8), confirming that the association is not explained by molecular subtype alone. A broader complete-case clinicopathologic model additionally adjusting for Nottingham grade and nodal stage was treated as supportive rather than primary, because 31% Nottingham-grade missingness reduced the analysable cohort to 611 subjects (49 events) and an events-per-variable ratio of approximately 5.4; in that reduced cohort the entropy association remained directionally consistent (HR = 0.530, 95% CI 0.297–0.948, p = 0.032; Supplementary Table S8.9). A Duke-internal median sensitivity analysis (threshold 2.0226) produced the expected balanced 454/454 split and a stronger, more precisely estimated association (log-rank p = 0.0007; HR = 0.435, 95% CI 0.269–0.703; Supplementary Fig. S12), demonstrating that the prognostic signal is robust to the specific numeric threshold. The absolute risk difference under the internal-median threshold (11.2% versus 5.5%) corresponds to a number needed to screen of approximately 18 patients to identify one additional distant recurrence event. Duke DRFS curves are shown in Fig. 8.

Subtype-stratified Cox regression yielded directionally consistent hazard ratios below 1.0 in all estimable major subtype groups, with no evidence of effect heterogeneity under either a categorical (interaction p = 0.833) or ordinal (interaction p = 0.672) specification (Supplementary Table S8.10; Fig. S14). In the pre-specified pCR secondary analysis (NAT subset n = 312), Entropy_Tumor showed null pCR discrimination (AUROC = 0.515, 95% CI 0.440–0.594; Supplementary Table S8.11), replicating the pCR–survival dissociation established in I-SPY2 (combined OR = 0.97, p = 0.926) and confirming that the entropy-defined structural phenotype functions as a recurrence-risk organiser rather than a pCR classifier. The coexistence of AUROC = 0.515 for pCR and HR = 0.466 for DRFS within the same Duke cohort constitutes the most direct within-cohort evidence for the pCR–survival dissociation.

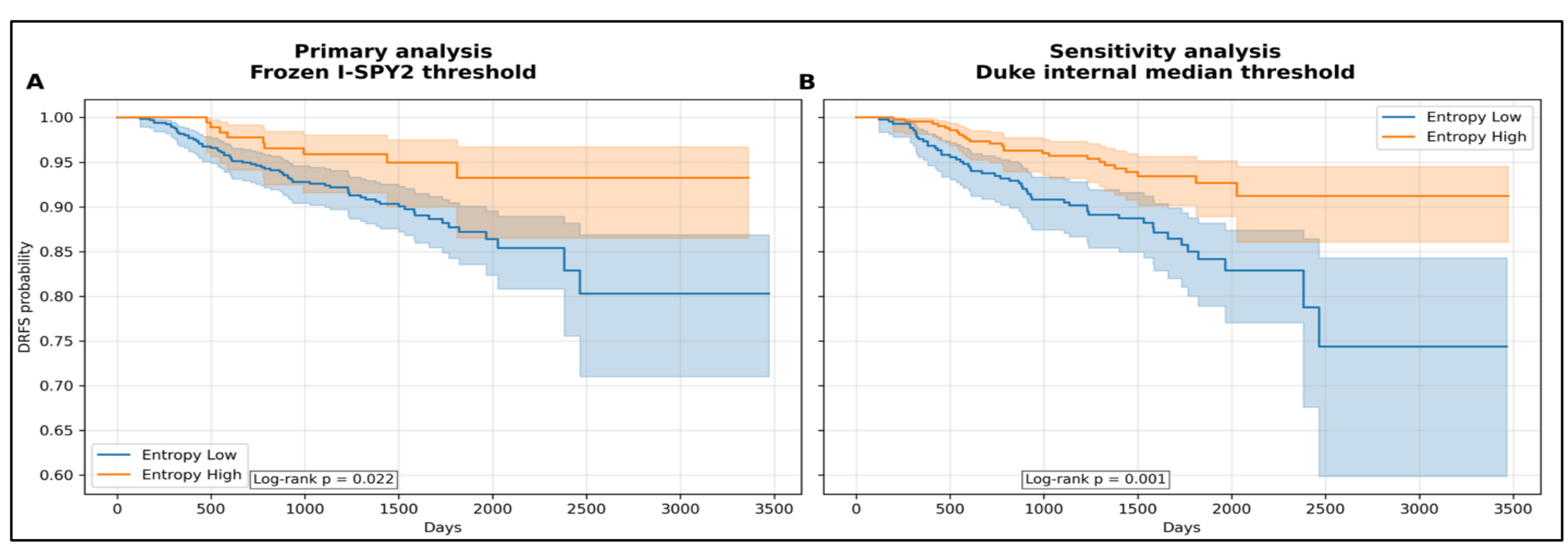

***Fig. 8.*** *Distant recurrence-free survival stratification by pretreatment entropy in the Duke external cohort (n = 908). (A) Primary analysis using the pre-specified I-SPY2 training-partition median entropy threshold (2.6500518); Entropy_low (n = 709) versus Entropy_high (n = 199); log-rank p = 0.022; unadjusted HR = 0.452 (95% CI 0.225–0.906). (B) Pre-specified sensitivity analysis using the Duke internal median threshold (2.0226); balanced 454/454 split (Entropy_low 51 events, Entropy_high 25 events); log-rank p = 0.0007; HR = 0.435 (95% CI 0.269–0.703). Entropy_high subjects experienced consistently lower hazard of distant recurrence under both thresholding schemes, demonstrating that the prognostic signal is not an artifact of the pre-specified I-SPY2 boundary. Shaded bands denote 95% confidence intervals.*

### 3.10. Cross-Cohort Transportability Synthesis Under the RFCA Framework

The three external cohorts resolve into a coherent transportability framework rather than three disconnected validation exercises, each occupying a distinct position in transportability space (Table 7).

**Table 7. Cross-Cohort Transportability Matrix Under the RFCA Framework**

| Cohort | Feature family | Stage 1 status | Support / geometry status | Key finding | Interpretation |
|---|---|---|---|---|---|
| UCSF | Tier-2 FTV | REGIME_COMPATIBLE (d = −0.039 to +0.030 for dynamic features) | 91.8% in-support; State 1 fully supported (median centroid distance = 0.474) | Log-rank p = 0.0042; RMST difference = 22.6 months (τ = 60 months); pCR flat across states (p = 1.000) | Geometry and prognostic transport — feature-family commensurability and manifold support both preserved |
| Duke | Entropy_Tumor | REGIME_COMPATIBLE (d = −0.020) | GEOMETRY_ENTANGLED (ρ = −0.729 with MajorAxis; bounding-box extraction) | primary adjusted HR = 0.466 (p = 0.0317); major-axis-adjusted HR = 0.839 (p = 0.642); residualized HR = 1.11 (p = 0.603); attenuation extends across entropy family | Regime-compatible but geometry-entangled — Stage 1 passage insufficient when extraction regime differs |
| I-SPY1 | Entropy (raw-DCE SER) | PARTIALLY_SHIFTED (d = 1.354 after forensic cleanup; entropy not projected) | — (Stage 1 failed; projection not performed) | d = 1.354 PARTIALLY_SHIFTED under all constructions; stored-map constructions failed at d = 2.57–4.78 SD | Non-transportable feature family — protocol lineage and label similarity insufficient to guarantee numeric commensurability |
| I-SPY1 | Tier-2 FTV | REGIME_COMPATIBLE (ΔFTV_10 d = −0.039; ΔFTV_20 d = 0.030) | 87.3% in-support overall; State 1 fully supported (100% within 95th-percentile; median distance = 0.541); States 0/2 peripheral (median distance > 2.7) | pCR flat across states ($\chi^2$ p = 0.956); centroid-distance Cox HR = 2.04 (95% CI 1.15–3.64, p = 0.015) within State 1; coarse binary survival null (log-rank p = 0.801) | Geometry transport without coarse prognostic replication — null binary survival explained by peripheral placement of States 0 and 2; recurrence microstructure detectable within fully supported state |

External radiomic transportability is not a binary cohort-level property; it is a conditional property of a feature family under a specific extraction and support geometry. UCSF provided the clearest demonstration of successful Tier-2 transport: when both feature-family commensurability and manifold support are preserved, external geometry transport extends to prognostic transport. Duke demonstrated the second-order condition — Stage 1 distributional commensurability is necessary but not sufficient when extraction geometries differ — and the RFCA's Stage 4 audit detected this boundary condition, preventing an invalid inference that standard validation would have missed. I-SPY1 provided the within-cohort logical proof that transportability is a representation-family property: in the same 126 subjects, entropy failed Stage 1 (d = 1.354) while FTV dynamics passed near-perfectly (d = 0.030). Together these four rows are more informative than uniformly positive external replications would have been: both success conditions and failure modes are mechanistically interpretable under the same framework, establishing that the auditing conditions are necessary and not merely sufficient.

The results establish two inseparable contributions: the identification of a structural phenotype that organises longitudinal treatment-response geometry independently of molecular classification and binary surgical endpoints, and a representation-family transportability framework that determines the conditions under which that phenotype can be externally validated. The biological finding motivates the methodological infrastructure; the methodological infrastructure is what makes the biological finding credible across cohorts. The following sections discuss each in order.

## 4. Discussion

### 4.1. Biological Interpretation of the Response Manifold

In this study, "structure" refers to the spatial complexity of tumour enhancement on pretreatment DCE-MRI, operationalised primarily by Shannon entropy, and should be interpreted as an imaging phenotype of architectural heterogeneity rather than a direct histologic mechanism. The central result of this study is a single, quantitatively sharp dissociation: the dominant axis of treatment-response geometry is unrecoverable from the full conventional clinical and genomic stack (cross-validated $R^2 = -0.023$) yet becomes strongly recoverable the moment baseline structural entropy is added ($R^2 = 0.876$; $\Delta R^2 = 0.899$). The analyses presented in Sections 3.1–3.8 constitute different tests of a single central claim: that pretreatment structural tumour organisation defines a biologically distinct, clinically consequential dimension of treatment response that the conventional clinical and genomic stack does not resolve. The manifold reveals a latent structural phenotype of treatment response — with architectural complexity as the dominant axis, tumour burden second, and treatment-induced structural reorganisation third — that is independent of the entire current molecular classification stack simultaneously. Two patients of identical molecular subtype who occupy different LDRS states differ primarily in baseline structural heterogeneity — a dimension that molecular profiling alone does not resolve. The structural phenotype is not merely uncorrelated with established molecular classifiers; it remains strongly separative within every stratum they define, including all four receptor-defined subtypes, all four PAM50 intrinsic subtypes, and both MammaPrint risk groups — demonstrating that current genomic and transcriptomic classifiers do not resolve this dimension of tumour organisation. This positions imaging-defined structural phenotyping not as a surrogate for molecular classification but as a biologically independent complement to it, consistent with evidence linking intratumoural architectural complexity to treatment resistance and stromal microenvironmental organisation [13, 14, 16]. This resolution limit is most directly visible within MammaPrint-High patients — the subgroup that genomic classification identifies as carrying the highest proliferative risk: even within this genomically homogeneous stratum, entropy separates two structurally distinct populations (Supplementary Section S2.7), confirming that the structural phenotype captures a dimension of tumour organisation that the most widely deployed genomic risk classifier does not access.

The DLRM validation establishes that this organisation is intrinsic to the data's covariance geometry rather than a post-hoc interpretation of generic variance components: imposing the biological axis structure as a construction constraint reproduces, at near-identity, the manifold that unconstrained PCA recovers independently (Supplementary Table S2.9). Had the entropy/burden/remodelling structure been incidental to variance maximisation, the two solutions would diverge. This formal test of intrinsicness is rarely made explicit in unsupervised radiomic manifold work, and the DLRM provides a reusable paradigm for manifold-based radiomic phenotyping. Although baseline entropy is the dominant scalar correlate of PC1, the manifold offers three capabilities entropy alone cannot: continuous multidimensional risk grading (centroid distance HR = 1.404, $p = 0.030$, independently significant alongside PC1), a secondary remodelling axis (PC3) orthogonal to baseline phenotype, and geometric support conditioning under RFCA, which requires manifold coordinates rather than a scalar threshold.

Beyond the baseline phenotype axis, the manifold contained a secondary remodelling dimension. PC3 (9.8% variance explained) was dominated by longitudinal entropy-change features (ΔEntropy_10 loading = 0.501, ΔEntropy_30 = 0.500, ΔEntropy_20 = 0.444), while baseline entropy contributed only weakly (loading = -0.100). In incremental models, PC3 added substantial explanatory value beyond PC1 for structural remodelling outcomes ($\Delta R^2 = 0.464$-$0.592$ for entropy-change targets, all $p < 10^{-54}$). Importantly, this effect persisted after residualisation against both baseline entropy and baseline tumour burden: PC3 alone explained 45-59% of the remaining variance in entropy-remodelling targets after full baseline adjustment, whereas PC1 explained only 1.5-1.8%. PC3 did not improve pCR discrimination ($p = 0.815$), indicating that it captures treatment-related structural reorganisation rather than endpoint biology. The manifold is therefore not one-dimensional: PC1 defines baseline structural phenotype (what the tumour is), whereas PC3 defines treatment-related structural remodelling (what the tumour does under therapy) (Supplementary Table S13.6).

The longitudinal trajectory analysis extends the baseline organisation into a dynamic claim. Baseline structural heterogeneity predicts not just the static architecture of the tumour at diagnosis but the character of treatment-induced structural reorganisation under therapy — entropy-defined groups diverge in reorganisation trajectories across all treatment intervals while volumetric trajectories converge completely by T1. This dissociation provides a mechanistic account of why structural phenotype stratifies survival independently of the binary surgical endpoint: if structural reorganisation diverges even among patients with identical outcomes, the biology that governs long-term recurrence may remain unresolved by pCR.

### 4.2. Persistence, Concealment, and the Information Crossover

The structural phenotype behaves as an enduring biological property rather than a transient response-state fluctuation: rank-order persistence through therapy ($\rho = 0.736$) shows that treatment does not systematically reorganise the structural hierarchy, even though neoadjuvant chemotherapy profoundly remodels tumour architecture. The clinical consequence is sharpened by the volumetric deception finding — within pCR achievers, the most structurally disordered tumours showed the strongest early volumetric shrinkage ($\rho = -0.265$), so the patients who appear most treatment-responsive by conventional imaging are the ones whose structural biology is most concerning. This produces an information crossover: by the first on-treatment assessment, volumetric measurement has lost discriminative power between phenotype groups while the structural axis retains maximal separation throughout treatment. Any framework relying exclusively on post-baseline volumetric response is operating beyond the point at which its discriminative capacity has expired.

The incremental pCR ladder reflects the hierarchical biology of treatment response: intrinsic tumour biology dominates, while structural heterogeneity and tumour burden add complementary information beyond molecular classification [13, 14]. The subtype-adjusted signal test (§3.3.2) sets the precise scope — adding manifold variables to an explicit molecular baseline produced no significant pCR improvement. This is not a failure but a clarification: the manifold characterises the structural response geometry whose primary clinical expression is long-term recurrence risk, not pCR reclassification once subtype is known.

The within-pCR dissociation reinforces this interpretation directly. Longitudinal trajectory geometry encodes biology that neither baseline snapshot nor surgical endpoint resolves — serial imaging is the source of the trajectory information the manifold organises, and baseline-only models cannot recover the treatment-induced structural reorganisation dimension captured by PC3 ($\rho = 0.679$ with $\Delta$Entropy_30). The performance ceiling of AUROC = 0.781 appears sample-size-limited rather than model-class-limited: nonlinear alternatives (XGBoost, RBF-SVM) fell below the elastic-net benchmark on the locked test set, and learning-curve analysis showed no plateau at full training size (terminal slope +0.030 AUROC per 20% additional data), indicating that exceeding the current ceiling will likely require larger training cohorts rather than greater tabular model flexibility (Supplementary Section S5.12).

### 4.3. The pCR-Survival Dissociation and Clinical Consequences

The phenotype's clinical expression is convergent rather than singular. Within-pCR trajectory divergence and volumetric deception show that standard response metrics can obscure structurally distinct reorganisation despite identical surgical outcomes (§3.3.4); within RFCA-validated support regions continuous manifold geometry carries graded recurrence risk in I-SPY1; pre-specified Tier-2 geometry reproduces absolute survival separation in UCSF; and the Duke association proves geometry-entangled under bounding-box extraction (§3.8.5). The UCSF result is the most stringent: pre-specified trajectory geometry retained significant prognostic stratification across institutions, scanner generations, and a two-decade acquisition gap (log-rank $p = 0.0042$; RMST advantage 22.6 months) without recalibration. That entropy had to be excluded from UCSF projection — despite using the same first-order feature and discretization — is itself a contribution: it documents a case where matching the extraction parameters was insufficient because the source-image representation differed (UCSF PE1-phase vs I-SPY2 SER-native), making the UCSF separation a conservative lower bound on what the complete manifold might achieve under representation-harmonised conditions.

I-SPY1 provided the clearest within-cohort demonstration that this success is not a cohort-level property. In the same 126 subjects, under the same PyRadiomics configuration and ACRIN acquisition lineage: entropy failed RFCA Stage 1 ($d = 1.354$) while FTV dynamics passed near-perfectly ($d = 0.030$) — proving that transportability is a representation-family property, not a cohort-level property. The null coarse survival result (log-rank $p = 0.801$) is mechanistically explained by the peripheral manifold placement of States 0 and 2 rather than absence of biological signal; within the fully supported State 1 subgroup, centroid distance independently predicted recurrence-free survival (HR = 2.04, 95% CI 1.15–3.64, $p = 0.015$), establishing that continuous manifold placement geometry encodes recurrence-relevant microstructure beyond coarse state assignment. I-SPY1 also independently replicates the pCR-trajectory dissociation across a structurally distinct acquisition era, reinforcing the paper's central biological argument in an external cohort with different acquisition parameters, scanner generation, and treatment structure. This result establishes that the latent structural phenotype encodes graded recurrence-relevant geometry within transported manifold states, not merely a dichotomous threshold distinction — a property visible only when external projection support is explicitly verified, and one that coarse binary survival comparisons would miss entirely. The binary entropy threshold used throughout this study is therefore a scientifically conservative operationalisation of what the data indicate is a continuous phenotype: the median split recovers a statistically tractable dichotomy, but the underlying manifold geometry likely encodes finer-grained risk structure than any single threshold can capture.

The Duke validation clarifies a boundary condition for the structural hypothesis. The persistence of the entropy-defined structural phenotype after adjustment for molecular subtype, Nottingham grade, and nodal stage in 908 subjects with 76 distant recurrence events establishes that pretreatment MRI structural heterogeneity captures a dimension of tumour organisation not fully represented by the clinical variables currently used for neoadjuvant risk stratification (primary adjusted HR = 0.466, 95% CI 0.232–0.935, $p = 0.0317$). The primary adjusted, subtype-adjusted, supporting complete-case, and internal-median analyses all converged on the same directional conclusion, establishing that the association is not a threshold artefact and is not explained by molecular subtype alone.

The coexistence of a null entropy–pCR association in both I-SPY2 (OR = 0.97, $p = 0.926$) and Duke (AUROC = 0.515) alongside a significant recurrence association in both cohorts is not a paradox but a precise characterisation of what entropy measures and what pCR misses — the paper's central conceptual finding. Pathologic complete response is a binary threshold on a volume endpoint that captures drug-sensitivity biology at the surgical window [46, 48, 51], whereas distant recurrence reflects cumulative residual-disease biology and dormancy dynamics long after treatment [47, 49, 65]. The within-pCR trajectory analysis (§3.3.4) gives the direct mechanistic account: among I-SPY2 pCR achievers ($n = 220$), entropy-defined groups followed structurally distinct longitudinal reorganisation at every timepoint (all $p < 10^{-19}$) while volumetric trajectories were indistinguishable (all $p > 0.069$). Structural reorganisation diverges even among patients with uniformly favourable surgical outcomes — a dimension of biology pCR does not resolve — and the Duke DRFS result is consistent with the same mechanism. The clinical implication is specific: pCR achievers are currently grouped by a uniformly favourable surgical outcome irrespective of pretreatment structural phenotype, despite evidence that recurrence biology differs even within pCR-positive populations [52–55]. The within-pCR structural divergence identifies a candidate basis for differentiating post-surgical surveillance intensity in precisely the subgroup where endpoint-based stratification offers the least biological resolution. The practical opportunity is therefore not immediate treatment selection, but a pretreatment, imaging-only handle for identifying structurally distinct response states that endpoint labels currently collapse. This requires prospective validation, not a new intervention — only representation-aware extraction, contour-based segmentation, and pre-specified adjustment for tumour size.

A companion study [69] is examining the molecular correlates and clinical consequences of the structural phenotype characterised here — questions the present analysis deliberately leaves open. Using a separate analytical framework, that ongoing work is evaluating whether the structural distinction corresponds to a distinct, PAM50-independent molecular programme, whether a computationally independent image representation recovers the same phenotype, and whether the same structural axis carries prognostic information within the pathologic-complete-response population across independent external cohorts. This companion work is cited only as convergent context: the discovery, molecular-independence, persistence, external-validation, and RFCA claims of the present study rest entirely on the analyses reported above and do not depend on that ongoing companion analysis.

One boundary condition must be stated precisely. The entropy hazard ratio attenuated to non-significance after adjustment for MRI-derived major-axis length while preserving the protective direction (HR = 0.839, 95% CI 0.401–1.756, $p = 0.642$), consistent with partial collinearity between bounding-box-derived entropy and lesion spatial extent rather than reversal of the primary association. This is a diagnostic success: the RFCA framework identified a boundary condition that standard validation would have reported as a positive result. Geometry-coupling analysis established the mechanism directly — Entropy_Tumor showed a strong negative correlation with imaging-derived major-axis length (ρ = −0.729) but no meaningful correlation with clinical tumour size (ρ = +0.022), establishing that the entanglement is a property of bounding-box extraction, not of architectural disorder as a construct. In I-SPY2, where contour-based segmentation separates structural heterogeneity from spatial extent, explicit FTV residualization preserved manifold axes at near-identity (PC1 r = 0.992; PC2 r = 0.988). Prospective validation requires contour-based or otherwise geometry-decoupled extraction — a solvable methodological condition, not a fundamental biological barrier.

The three external cohorts resolve into a coherent transportability framework rather than three disconnected results. UCSF demonstrated full geometry and prognosis transport when representation-family commensurability and manifold support are both preserved. I-SPY1 proved within a single cohort that transportability is a representation-family property that cannot be inferred from protocol lineage or surface similarity. Duke demonstrated the second-order condition: Stage 1 distributional commensurability is necessary but not sufficient when extraction geometries differ — the RFCA's Stage 4 audit detected this, preventing an invalid inference that standard validation would have missed. External radiomic transportability is therefore not a binary cohort-level property but a conditional property of a feature family under a specific extraction and support geometry.

Together, I-SPY1 distributional shift and Duke geometry entanglement define a two-level commensurability taxonomy that the RFCA framework addresses (§3.10; Supplementary Table S13.1). The primary trajectory and survival design comprises four cohorts (Table 1). Transportability is therefore conditional across extraction pathways, and neither failure mode subsumes the other.

This validation also spans an unusual acquisition range: the four cohorts cover three decades of treatment and imaging, from single-centre studies to multi-vendor mixed-field-strength environments and from research-protocol segmentations to precomputed bounding-box matrices. That a single pre-specified manifold trained on I-SPY2 recovers prognostically meaningful states in UCSF (log-rank p = 0.0042) and recurrence-relevant microstructure in I-SPY1 (within-State-1 HR = 2.04, p = 0.015) without cross-era recalibration is a stringent test of phenotype stability that few radiomic biomarkers have undergone.

#### 4.4. Support-Conditioned Clinical Meaning

Having established that the phenotype exists, persists, and is obscured within pCR, the remaining question is whether its continuous geometry carries graded clinical meaning — and under what conditions. The support-conditioned continuous risk analysis answers both questions simultaneously. In the I-SPY1 projected cohort, continuous PC1 showed no significant association with recurrence-free survival across all projected subjects (HR = 0.805, p = 0.135). After restriction to the RFCA-validated support region, the same axis became significantly associated with outcome (HR = 0.624, 95% CI 0.453–0.861, p = 0.004). This is not a post-hoc subgroup analysis: the support boundary was defined by the training manifold geometry before any I-SPY1 outcome data were accessed, and the out-of-support subjects occupied a genuinely different geometric regime (Cohen's d up to 3.65). Within the supported region, both axial position (PC1) and geometric deviation (centroid distance, HR = 1.404, p = 0.030) carried outcome-relevant information, demonstrating that the manifold encodes multidimensional graded risk structure, not merely a binary threshold. The phenotype is real, durable, hidden by conventional metrics, and its continuous geometry carries clinically meaningful information exactly where the representation-family framework says the inputs are valid — and not where they are not. RFCA does not merely audit past analyses; it makes a prospective prediction about where manifold geometry can be trusted, and the survival data confirm that prediction.

#### 4.5. Representation-Family Transportability as a Methodological Framework

Section 4.4 demonstrated that manifold geometry does not carry uniform clinical meaning across all projected subjects; it becomes interpretable only where representation-family support conditions are satisfied. The broader implication is that external radiomic validation must be governed by commensurability verification before performance claims are read as biological replication. Existing radiomic validation standards implicitly assume that feature labels define a portable representation. This study shows that assumption is false: feature labels can occupy incommensurable numeric regimes across cohorts, and distributional commensurability can coexist with geometry entanglement that renders the signal biologically uninterpretable, as in Duke where the entropy association attenuates to non-significance after adjustment for lesion major-axis length (§S8.11–S8.18). RFCA operationalises the correction: it asks not whether a named feature transfers across cohorts, but whether the representation family supports transfer at all — a prior question that feature-label matching alone cannot answer. The practical consequence is methodological: any radiomic biomarker study reporting cross-institutional generalisation without first establishing representation-family commensurability is making an inferential claim its own data cannot support, and such discrimination should not be read as biological replication. The complete four-stage decision procedure is illustrated in Supplementary Fig. S7.1.

### 4.6. Relationship to Prior Work

The DCE-MRI radiomics literature in neoadjuvant breast cancer has emphasized pCR discrimination across increasingly large cohorts, with fewer studies modelling long-term survival and fewer still evaluating whether imaging-derived features transport across acquisition environments [42–45]. A structured comparison against the most directly relevant studies is provided in Supplementary Table S12.1.

The present work differs from this literature in five structural respects. The primary claim is biological characterisation of unsupervised response space rather than pCR discrimination optimisation — a distinction that reframes the paper's evaluation criteria. The evaluation framework combines partition-locked inference, nested cross-validation, bootstrap inference, label permutation, and calibration sensitivity under a single reproducibility contract that individually standard practices rarely achieve together. External survival validation across three independent cohorts under distinct inference protocols constitutes a substantially stronger generalisability test than held-out pCR discrimination from a single acquisition environment. The study introduces a pre-specified representation-family transportability audit as a necessary step prior to cross-cohort inference, demonstrating empirically that transportability is a feature-family property requiring explicit distributional verification. And the Stage 4 extraction-geometry audit establishes that distributional commensurability alone is insufficient for biological inference when extraction geometries differ — a second-order condition not previously formalised in radiomic validation frameworks.

The most directly complementary prior work is Chitalia et al. [14], which identified early heterogeneity phenotypes associated with recurrence-free survival within I-SPY1; the present study extends that biological hypothesis across three additional independent cohorts under locked inference, and with matched imaging–transcriptomic confirmation of molecular independence at the PAM50 level. Su et al. [45] provides the closest parallel in cohort scale and unsupervised manifold design but is limited to baseline pretreatment features without longitudinal trajectory encoding and without representation-family transportability assessment.

The framework's anchoring on FTV and Shannon entropy — each mapping to a defined biological construct — proved decisive: the BHI composite failed commensurability in Duke, where only one of eleven features transported, so a framework built on pipeline-specific high-order texture would have had no transportable core. The I-SPY1 within-cohort contrast, where entropy failed Stage 1 while FTV dynamics passed in the same subjects, shows this is a representation-family property rather than a cohort-level one, making anchor-feature selection a prerequisite for cross-cohort radiomic inference rather than a modelling preference.

### 4.7. Limitations

This study has several limitations, set largely by the scope of publicly available data. The external cohorts are retrospective and public, and no single dataset provides multi-timepoint expert segmentations, long-term recurrence outcomes, and harmonised extraction together. The UCSF validation is small (49 subjects, 15 events), so the adjusted Cox model was underpowered (HR = 1.44, 95% CI 0.76–2.73); the external claim therefore rests on the convergent log-rank separation ($p = 0.0042$) and RMST advantage (22.6 months), not on the adjusted estimate. Acquisition era differs markedly across UCSF, I-SPY1, and I-SPY2.

The Duke validation carries the strongest boundary conditions: bounding-box segmentation entangles textural disorder with tumour spatial extent (§3.8.5), only a single pretreatment timepoint is available — limiting Duke to the baseline heterogeneity axis — and complete-case analysis reduced the cohort from 908 to 611 subjects under 31% Nottingham-grade missingness whose mechanism cannot be confirmed. The RFCA framework, though demonstrated across multiple operational failure modes here, is proof-of-concept: its validity across other biomarker classes, diseases, and modalities, and the calibration of its operational thresholds, remain to be established prospectively. The PC3 remodelling axis is not yet validated against external outcomes. All boundary conditions are reported transparently and are mechanistically interpretable; each is solvable by prospective, multi-institutional validation with contour-based segmentation, harmonised extraction, longitudinal imaging, and pre-specified representation-family auditing.

### 4.8. Future Directions

Prospective multi-institutional neoadjuvant datasets with longitudinal DCE-MRI, harmonised contour-based segmentation, and pre-specified entropy and FTV extraction represent the most direct path to exceeding the current performance ceiling and decoupling entropy from extraction-geometry confounds. The molecular substrate of the entropy-defined architectural phenotype remains to be fully established; spatial transcriptomics or matched computational pathology could reveal cellular-level correlates without changing the present finding of molecular orthogonality. The pCR-survival dissociation, replicated cross-cohort in both I-SPY2 and Duke, motivates prospective evaluation as a recurrence-risk organiser informing post-surgical surveillance intensity. Whether the PC3 remodelling axis carries independent prognostic value beyond baseline phenotype is testable in cohorts with serial imaging and long-term follow-up.

## Conclusion

Unsupervised manifold learning applied to serial DCE-MRI trajectories revealed a latent structural phenotype of treatment response — with baseline entropy as its dominant scalar correlate and treatment-induced structural remodelling as an orthogonal second axis (PC3) — that is intrinsic to the data covariance geometry (DLRM convergence at $|r| \geq 0.952$; ARI = 1.000) and independent of the full conventional clinical and genomic classification stack.

Five convergent lines of evidence establish this phenotype as a distinct, durable, and clinically consequential structural dimension. First, the dominant manifold axis was not recoverable from tumour burden, age, receptor subtype, MammaPrint, treatment arm, and PAM50 combined (cross-validated $R^2$ = -0.023), but became highly recoverable after entropy addition (cross-validated $R^2$ = 0.876) — the phenotype is invisible to the established classification architecture. Second, the structural phenotype persisted through therapy (88.2% state retention; continuous $\rho = 0.736$), establishing it as a durable biological property rather than a transient response state. Third, within pCR achievers, structurally disordered tumours showed stronger rather than weaker early volumetric shrinkage ($\rho = -0.265$; size-adjusted $\rho = -0.262$) yet remained structurally disordered, meaning standard response metrics can obscure persistent structural disorder. Fourth, continuous manifold geometry became prognostically significant only within RFCA-supported regions (in-support HR = 0.624, $p = 0.004$ versus non-significant overall), demonstrating that the four-stage Representation-Family Commensurability Assessment correctly predicts where manifold geometry carries clinical meaning. Fifth, the RFCA framework identified two distinct commensurability failure modes across the external cohorts — cross-cohort distributional shift and extraction-geometry entanglement — providing reusable infrastructure for representation-aware radiomic validation beyond this study.

The latent structural phenotype identified here is not a surrogate for the conventional clinical and genomic stack; that stack is blind to it. It is not transient under therapy; its ordering persists. It is not faithfully represented by standard volumetric response language; that language can obscure it. And it is not clinically interpretable everywhere a projection can be computed; its continuous geometry carries prognostic meaning only where representation-family support conditions are satisfied. The consequence is methodological as well as biological: radiomic validation can no longer rely on feature-label matching alone, because without representation-family commensurability auditing, biological replication cannot be distinguished from numerical coincidence. Together, these findings establish pretreatment DCE-MRI structural entropy as the dominant scalar correlate of a latent response phenotype that is molecularly independent, persistent through therapy, externally recurrence-relevant under representation-aware validation, and incompletely captured by pCR or volumetric response language.

## Data and Code Availability

The I-SPY2 and UCSF imaging datasets are available through The Cancer Imaging Archive (TCIA) [18, 56], subject to TCIA data access policies. The I-SPY1 dataset is available through TCIA [63]. Duke Breast Cancer MRI is also publicly available through TCIA [57], subject to its corresponding access and usage terms. All preprocessing scripts, model training code, locked parameter artifacts (including scalers, PCA transforms, cluster centroids, elastic-net coefficients, calibration models, and decision thresholds), and figure-reproduction code will be made publicly available in a dedicated repository upon acceptance.
Supplementary materials and reproducibility documentation for this arXiv preprint are available at: https://drive.google.com/file/d/197DqCiKKXZz83mT0hnPuK2dGoJMV7QB5/view

## Acknowledgements

The authors acknowledge the investigators of the I-SPY 1, I-SPY 2, ACRIN, UCSF, and Duke imaging trials for making their data publicly available through The Cancer Imaging Archive.

## Declaration of competing interests

The authors declare no competing interests.

## Funding

This research did not receive any specific grant from funding agencies in the public, commercial, or not-for-profit sectors.

## Credit author statement

Dattatreya Kantha: Conceptualization, Methodology, Software, Formal analysis, Investigation, Data curation, Visualization, Writing – original draft. Murray H. Loew: Conceptualization, Supervision, Writing – review and editing.